\documentclass[sigplan,screen,table]{acmart}
\renewcommand\footnotetextcopyrightpermission[1]{}
\usepackage{xcolor}
\usepackage{mathptmx}
\usepackage[subtle]{savetrees}

\usepackage{tikz}
\usepackage{amsmath}
\usepackage{titlesec}

\usepackage{tabu}
\usepackage{listofitems}
\usepackage{xstring}
\usepackage{ifthen}
\usepackage{tikz} %
\usetikzlibrary{arrows.meta,bending,intersections,decorations.markings,shapes.geometric,decorations.pathreplacing,calc, trees, fadings, shapes.multipart}
\usetikzlibrary{fit,patterns,plotmarks,backgrounds}
\usetikzlibrary{positioning}

\usepackage{booktabs}

\usepackage{collcell}
\usepackage{float}
\usepackage{graphicx}
\usepackage[utf8]{inputenc}
\usepackage{multirow}
\usepackage[group-separator={,},per-mode=symbol]{siunitx}
\usepackage{subcaption}
\usepackage{xspace}

\usepackage{wrapfig}

\usepackage{enumitem}
\setlist[itemize]{noitemsep,topsep=0pt,after=\vspace{.5\baselineskip}}
\setlist[enumerate]{noitemsep,topsep=0pt,after=\vspace{.5\baselineskip}}

\usepackage{hyperref}
\usepackage[nameinlink,noabbrev]{cleveref}
\usepackage{dsfont}
\usepackage{relsize} %

\usepackage[ruled,vlined]{algorithm2e}
\usepackage{natbib}
\usepackage[normalem]{ulem}

\def\Snospace~{\S{}}

\crefformat{figure}{Fig.~#2#1#3}
\crefformat{section}{\S#2#1#3}
\crefformat{table}{Table~#2#1#3}
\crefmultiformat{section}{\S\S#2#1#3}{ and~#2#1#3}{, #2#1#3}{, and~#2#1#3}

\usepackage{setspace}  %
\newif\ifEditMode

\EditModetrue
\EditModefalse

\hypersetup{pdfstartview=FitH,
	pdfpagelayout=SinglePage,
	bookmarksnumbered=true,
	bookmarksopen=true,
	colorlinks=true,
	citecolor=blue,
	linkcolor= blue,
	urlcolor=blue}

\newcommand{\parab}[1]{\vspace{0.02in}\noindent{\textbf{#1}}\quad}

\newcommand{\figcap}[1]{\caption{\textit{#1}}}
\newcommand{\sfigcap}[1]{\caption{\textit{\small #1}}}

\newcommand{\thead}[1]{\textbf{\textit{#1}}}
\newcommand{\ums}[1]{\SI{#1}{\milli\second}}

\newcommand{\sys}{Groundhog\xspace}
\newcommand{\manager}{\textit{manager}\xspace} %
\newcommand{\ow}{OpenWhisk\xspace}
\newcommand{\faasm}{{\sc{Faasm}}\xspace}

\newcommand{\wasm}{WebAssembly}
\newcommand{\Wasm}{\wasm}
\newcommand{\nodejs}{Node.js\xspace}

\newcommand{\eg}{e.g., \@}

\newcommand{\numcircle}[1]{
	\setbox0=\hbox{#1}%
	\dimen0\wd0%
	\divide\dimen0 by 2%
	\begin{tikzpicture}[baseline=(a.base)]%
		\useasboundingbox (-\the\dimen0,0pt) rectangle (\the\dimen0,1pt);
		\node[circle,draw,outer sep=0pt,inner sep=0.1ex] (a) {\textbf{{#1}}};
	\end{tikzpicture}
}

\newlength{\onecolgrid}
\newlength{\twocolgrid}
\newlength{\threecolgrid}
\newlength{\fourcolgrid}
\ifEditMode

\newcommand{\editlabel}[1]{\raisebox{.35ex}{\tiny \scshape[#1]}}

\xspace{}

\newcommand{\PENDING}[1]{
	\textcolor[HTML]{e67e22}{$\cdots$ \editlabel{Pending} \textit{#1} $\cdots$}\xspace{}}

\newcommand{\TODO}[1]{\textcolor[HTML]{e41a1c}{{#1}}}

\newcommand{\FIXME}[2]{
	\textcolor[HTML]{c0392b}{\editlabel{FIXME(#1)} \textbf{#2}}}
\newcommand{\MISSING}[1]{\textcolor[HTML]{de2d26}{\textbf{#1}}}

\newcommand{\REMOVE}[1]{
	{\small \textcolor[HTML]{a65628}{#1\\--- Consider deleting.}}}

\newcommand{\pd}[1]{\textcolor{magenta}{\relsize{-1}{\textbf{PD:} #1}}}
\newcommand{\dg}[1]{\textcolor{magenta}{{\relsize{-1}\textbf{DG:} #1}}}
\newcommand{\jon}[1]{\textcolor{magenta}{\relsize{-1}{\textbf{Jon:} #1}}}
\newcommand{\ma}[1]{\textcolor{magenta}{\relsize{-1}{\textbf{MA:} #1}}}
\newcommand{\update}[1]{{\color{blue}#1}}
\newcommand{\resolvepeter}[1]{{\color{violet}#1}}
\newcommand{\finalpass}[1]{{\color{teal}#1}}

\else
\makeatletter

\newcommand{\PENDING}[1]{\@bsphack\@esphack}
\newcommand{\TODO}[1]{\@bsphack\@esphack}
\newcommand{\FIXME}[2]{\@bsphack\@esphack}
\newcommand{\MISSING}[1]{\@bsphack\@esphack}
\newcommand{\REMOVE}[1]{\@bsphack\@esphack}

\newcommand{\pd}[1]{\@bsphack\@esphack}
\newcommand{\dg}[1]{\@bsphack\@esphack}
\newcommand{\jon}[1]{\@bsphack\@esphack}
\newcommand{\ma}[1]{\@bsphack\@esphack}
\newcommand{\update}[1]{#1}
\newcommand{\resolvepeter}[1]{#1}
\newcommand{\finalpass}[1]{#1}
\makeatother
\fi

\graphicspath{{./figures/}}

\titlespacing\section{0pt}{2 pt plus 1pt minus 2pt}{2 pt plus 1pt minus 2pt}
\titlespacing\subsection{0pt}{3 pt plus 1pt minus 2pt}{2 pt plus 1pt minus 2pt}
\titlespacing\subsubsection{0pt}{3 pt plus 1pt minus 2pt}{2 pt plus 1pt minus 2pt}
\setlist[itemize]{leftmargin=*}

\begin{document}

\acmConference{preprint}{arXiv}{2022}

\date{}
\title{\sys: Efficient Request Isolation in FaaS}
\settopmatter{authorsperrow=1} %

\author{Mohamed Alzayat\hspace{1.5em}Jonathan Mace\hspace{1.5em}Peter Druschel\hspace{1.5em}Deepak Garg}
\email{{alzayat,jcmace,druschel,dg}@mpi-sws.org}

\affiliation{%
	\institution{Max Planck Institute for Software Systems (MPI-SWS)}
	\streetaddress{Campus E1 5}
	\city{Saarbruecken}
	\state{Saarland}
	\country{Germany}
	\postcode{D-66123}
}

\renewcommand{\shortauthors}{M. Alzayat, J. Mace, P. Druschel, and D. Garg}

\begin{abstract}
	Security is a core responsibility for Function-as-a-Service (FaaS)
providers. The prevailing approach has each function execute in its
own container to isolate concurrent executions of different functions.
However, successive invocations of the same function commonly reuse
the runtime state of a previous invocation in order to avoid container
cold-start delays when invoking a function.  Although efficient, this
container reuse has security implications for functions that are
invoked on behalf of differently privileged users or administrative
domains: bugs in a function's implementation, third-party library, or
the language runtime \update{may leak private data from one invocation
of the function to subsequent invocations of the same function.}

\sys isolates sequential invocations of a function by efficiently
reverting to a clean state, free from any private data, after each
invocation.  The system exploits two properties of typical FaaS
platforms: each container executes at most one function at a time and
legitimate functions do not retain state across invocations.  This
enables {\sys} to efficiently snapshot and restore function state
between invocations in a manner that is independent of the programming
language/runtime and does not require any changes to existing
functions, libraries, language runtimes, or OS kernels. We describe
the design of \sys and its implementation in \ow, a popular \update{production-grade} open-source FaaS framework. On three existing
benchmark \update{suites}, {\sys} isolates sequential invocations with
modest overhead on end-to-end latency (median: 1.5\%, 95p: 7\%) and
throughput (median: 2.5\%, 95p: 49.6\%), \update{relative to an insecure baseline that reuses the container and runtime state.}

\end{abstract}
\maketitle

\section{Introduction}
\label{sec:intro}
Function-as-a-Service (FaaS) is an emerging high-level abstraction for
cloud applications. Tenants state their application logic as a
function implementation written in a high-level language like Python
or JavaScript. The FaaS provider in turn exports an HTTP/S endpoint,
which can be used to invoke the function with arguments and receive
results.  The FaaS provider is responsible for deploying and executing
the tenants' functions, provisioning and scaling resources as
workload demand fluctuates, and maintaining and multiplexing the
hardware and software infrastructure across different tenants and
functions.  FaaS has an `on-demand' payment model: a tenant only pays
for the compute time used to execute their functions.

\update{Among the core responsibilities of a FaaS provider is security. For scalability and efficiency, FaaS platforms multiplex functions of different tenants concurrently on a large pool of shared resources. FaaS platforms rely on various language-, process-, and VMM-based isolation mechanisms to isolate functions from one another: each function executes within its own execution environment and \emph{different functions} do not share the same execution environment. {\em Container isolation} is a commonly-used, general, low-entry-barrier function isolation mechanism that relies on standard OS process-isolation primitives}.
\update{An alternative to container isolation is VMM-based isolation,
  where each function executes in a separate VM.} Both container and
VMM-based isolation prevent a malicious or compromised function from
affecting the availability, integrity, and confidentiality of other
functions.

\update{So far, the focus of security in FaaS has been on isolating
  \emph{different functions} from each other.}  However, ideally, a
FaaS platform should provide the same degree of isolation among {\em
  sequential activations} of the \emph{same function}; otherwise, bugs
in a function implementation --- or a third-party library / runtime it
depends on --- may cause a leak of information from one activation of
a function to a subsequent one. This {\em sequential request
  isolation} is critical if a function can be invoked by, \update{or
  on behalf of,} differently privileged callers, such as unrelated
end-clients of a service built on top of the function. \update{For
  example, if the same function container is first invoked to service
  Alice's request and then invoked again to service Bob's request,
  there is a possibility that a bug in the function, some library or
  the language runtime causes some of Alice's data from the first
  request to be retained and later leaked into the response returned
  to Bob.} \update{It is this sequential request isolation problem
  that we focus on in this paper.}

\resolvepeter{A trivial way to attain sequential request isolation
  would be to run not just every function but also \emph{every
    activation of a function} in a freshly initialized
  container. However, this solution is problematic from the
  perspective of performance: Container initialization overheads are
  high, ranging from a few seconds when done naively to hundreds of
  milliseconds with existing solutions to reduce the cost of container
  cold-starts~\cite{coldstart-aws,coldstart-google,oakes2018sock,Stenbom2019refunction,catalyzer,REAP,faasnap,wang2019replayable},}
\update{which is higher than the basic execution time of a signficant
  fraction of FaaS functions (e.g., \cite{shahrad2020serverless}
  report function execution times in Microsoft Azure, with a median
  of~\ums{900} and a 25th \%-ile of~\ums{100} and we observe even
  lower function execution times in our experimental setup). Hence,
  this trivial solution to sequential request isolation would impose
  impractical overhead.}

\update{
This paper presents \sys, a system \update{that adds} lightweight
sequential request isolation to a FaaS framework that already uses containers to isolate different functions. Importantly, \sys reuses containers across requests to the same function, thus avoiding the per-activation container re-initialization cost of the trivial solution above}.\footnote{\resolvepeter{We describe our work in the context of FaaS
  platforms that already use containers to isolate different functions
  from each other, but similar design principles should apply to
  VMM-based isolation.}}  \sys is independent of the language, runtime,
or libraries used to implement functions, does not require changes to
function implementations, OS kernels or hypervisors, and preserves
most of the performance benefits of container reuse.  To the best of
our knowledge, \sys is the first system to do so.

\sys exploits two properties of FaaS platforms to enable a
general-purpose, lightweight, performant solution: (1) At most one function activation executes at any time in a container; and
(2) functions are not expected to retain runtime state across
activations.  Accordingly, the core of \sys's sequential request isolation is a
general, in-memory, {\em lightweight process snapshot/restore}
mechanism. \sys encapsulates each function in a \update{(containerized)} process, and takes a
snapshot of each function process' fully initialized state, just
before the function is being invoked for the first time. While this
state typically includes a fully initialized language runtime
including multiple threads, the function has not yet received
activation-specific arguments or credentials, and its state is
therefore guaranteed to be free of secrets. Subsequently, whenever the
function has finished an activation and returned its results, \sys
restores the function's process to the clean state recorded in the
snapshot.

\sys is secure because the restoration ensures that no data can leak from one activation to a
subsequent one.  \sys is efficient because the cost of restoring state
is roughly proportional to the amount of memory modified during an
activation. As we will show, most function activations modify only a
small proportion of the function process' total state. Finally, \sys
restores state {\em between activations of a function}, and therefore
does not contribute to a function's activation latency under low to
medium server load.

We have implemented \sys in C using commodity Linux kernel features.
We evaluate \sys in \ow using Python, \nodejs, and C functions
from the FaaSProfiler benchmark~\cite{faasProfiler}, pyperformance
\cite{pyperf} and PolyBench \cite{polybench} benchmarks, which cover a
wide variety of use cases. We demonstrate that \sys achieves sequential request
isolation with modest overhead on end-to-end latency (median: 1.5\%,
95p: 7\%) and throughput (median: 2.5\%, 95p: 49.6\%) \update{relative to an insecure baseline that reuses containers and runtimes}.
The main contributions of this paper include:\\
\numcircle{1} The design of a language and runtime-independent, in-memory
  \emph{lightweight process snapshot/restore} mechanism for achieving
  general-purpose sequential request isolation in FaaS while retaining the
  performance benefits of container reuse.\\
\numcircle{2} The design and implementation of \sys\footnote{\finalpass{The implementation will be open-sourced after the paper's publication.}}, a system that provides
  lightweight sequential request isolation using commodity Linux
  kernel features, and its integration into the \ow FaaS
  platform. \sys can be retrofitted to existing commercial systems
  without requiring any changes to existing functions, libraries,
  language runtimes, or OS kernels.\\
\numcircle{3} An experimental evaluation of \sys on functions from the FaaSProfiler, pyperformance, and PolyBench benchmarks within the \ow FaaS platform, which demonstrates that \sys provides sequential request isolation with low to modest overhead on function latency and peak throughput.

\section{Background}
\label{sec:motivation}
\parab{Functions and Requests} In the Function-as-a-Service (FaaS)
model, tenants upload \emph{functions} for execution by the cloud
provider.  A function is usually written in a high-level language,
accepts input arguments, and returns results.  The FaaS platform
exposes an HTTP/S endpoint to which the tenant's applications can send
\emph{requests} with arguments, and receive results in response.

\parab{Containers and Function Invocation} The FaaS provider is
responsible for executing functions on demand.  FaaS platforms execute
functions within \emph{containers}, which may take the form of
language-enforced compartments~\cite{cloudflare,faastly,faasm},
processes~\cite{akkus2018sand}, OS
containers~\cite{azure-functions,ow,openfaas,google-functions}, or
virtual machines~\cite{aws-lambda,firecracker}.
When a request arrives for a particular function, an instance of the
container needs to execute in order to serve the request.  The FaaS
platform may either instantiate a new container instance for the
requested function from scratch---a \emph{cold-start}---or
\emph{reuse} an existing idle container instance for the function if
one exists.

\parab{FaaS platform services} Many FaaS platforms offer tenant's
function implementations access to platform services. These services
include storage, such as file access to scratch storage on a local
disk, persistent key-value stores, or full relational database
backends. Platform services may also provide automatic invocation of
tenant's functions triggered by timers, writes to certain key-value
tuples, or updates to certain rows in a database.

\parab{Access control} FaaS providers support client authentication on
HTTP/S endpoints and minimally check if a caller is authorized to
invoke the function, based on an access control list provided by the
tenant.

Access to platform services by the function is controlled in this case
on a per-tenant basis.  Some FaaS providers like AWS-lambda, Azure
FunctionApps, Google Cloud Functions, and IBM Cloud
Functions~\cite{aws-lambda,azure-functions,google-functions,ibm-functions}
associate more fine-grained, per-caller\footnote{In this paper, the
	caller is the entity causally responsible for the activation of a
	function.} credentials to a function activation. Here, activations of
the same function can have access to different platform services
depending on the caller. Tenants can use this facility to control
information flow via platform services among differently privileged
callers of the same function, such as the different end-users of a tenant's
deployed application.

\parab{Security vs. Performance}
Security is a chief concern -- beyond per-tenant or per-caller access
control to functions and platform services, FaaS platforms must
prevent a buggy or malicious function from compromising other
functions or obtaining unauthorized access to platform services.
Containers are the key design choice for achieving such \emph{function
	isolation}.  Each container instance executes a single function.
Moreover, a container may be reused for repeated invocations of the
same function.

\parab{Sequential request isolation} Our focus in this work is
\emph{sequential request isolation}, \update{which isolates repeated
	invocations of the same function within the same container from each
	other. This isolation is important because bugs or compromises in a
	function implementation, or a third-party library or runtime the
	function relies on can cause confidentiality breaches by either (i)
	exposing private arguments of an activation to a subsequent
	activation of the function or (ii) using the credentials of an
	activation to obtain information from platform services and leak
	them to a subsequent, less privileged activation.
	
	\resolvepeter{A trivial method of sequential request isolation is to
		start each request in a freshly-initialized container (forcing a
		cold-start on each request). However, container initialization is
		expensive, as is well-known from studies on FaaS cold start
		latencies. Despite excellent progress on reducing container
		initialization
		costs~\cite{catalyzer,REAP,faasnap,wang2019replayable}, black-box
		techniques could still add hundreds of milliseconds of overhead on
		the critical execution path relative to standard insecure
		warm-container reuse. This overhead is of the same order of magnitude as a significant fraction of FaaS functions.}
	
	Consequently, we seek a different request isolation technique for FaaS
	that does does not rely on container cold-start on every request and
	has minimal performance impact relative to an insecure baseline that
	provides no isolation between sequential requests to the same
	function. Our solution adds only a few
	milliseconds of overhead off the critical path of a request (median:
	\ums{3.7}, 95p: \ums{16.1})
	and a minimal overhead for
	tracking modifications on the critical path (median: 1.5\%, 95p:
	7\%) relative to an insecure baseline that does not provide sequential request isolation.
We aim for a practical technique and particularly target a
\emph{black-box} approach that can be applied directly to functions
independent of language or FaaS runtime.}

\section{Design Preliminaries}
\label{sec:design}

\sys operates at the level of OS processes. It can be readily
integrated into FaaS platforms that encapsulate language runtimes and
functions in processes or containers, which includes most major FaaS platforms
currently in production use as far as we know. Moreover, \sys places
no restrictions on function implementations or the programming
language, runtime, and third-party libraries they rely on.  \sys
transparently interposes on API calls between a function
implementation and the FaaS platform. Function implementations as well
as the existing FaaS platform remain unchanged.

By interposing on a function's API calls, \sys detects when the
function is invoked and when its execution finishes and returns
results. \sys uses this information to transparently create an initial
snapshot of a newly created process and reverts its state after it has
finished executing an invocation. For this purpose, \sys relies on a
custom in-memory process snapshot/restore facility. The facility
relies on standard Linux functionality, such as soft-dirty bits to
track modified pages, the {\tt /proc} filesystem to monitor changes to
the process' address space mappings and read/write process memory, and
{\tt ptrace} to orchestrate state snapshot and restore.

\begin{figure}[tb]
\centering
\includegraphics[clip, trim=0.6cm 8cm 0.6cm 9cm,width={.48\textwidth}]{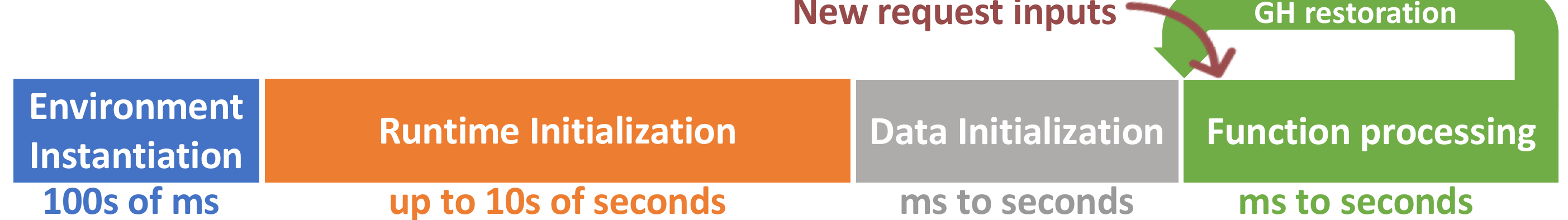}
\figcap{\sys container life cycle}
\label{fig:request-coldstart}
\end{figure}

\autoref{fig:request-coldstart} illustrates a function process's life
cycle when \sys is being used.  \sys avoids container, runtime, and
data initialization steps when reusing a function container (process),
and reverts the process' state in a median of \ums{3.7}  (10p: \ums{0.7}, 25p: \ums{1}, 75p: \ums{5.4}, 90p:
\ums{13}).
From the perspective of the FaaS platform, the
\sys-enabled container enjoys the benefits of container reuse, while
ensuring sequential request isolation irrespective of bugs in a
function's implementation, libraries, or runtime.

\subsection{Insights}
\label{sec:design:insights}
\sys relies on two properties of the FaaS model as implemented by
major FaaS platforms.

\parab{One-at-a-time function execution} In FaaS platforms, each
function container executes at most one request at a time.  For
scalable throughput, platforms create separate containers to
concurrently execute activations of different functions or multiple
activations of the same function.

\parab{Stateless functions} In the FaaS programming model, a function
implementation cannot expect that its internal state is retained
across activations.  To maintain persistent state, functions must
instead rely on external or platform services such as a key-value
store or a database backend.

Some FaaS platforms support {\em global state}, which can
be initialized using a tenant-supplied script that is executed when a
function container is initialized. \resolvepeter{Such an initialization step serves as mechanism to do operations or cache data that can be utilized by several activations regardless of their inputs (\eg populating data structures, download machine learning models).} This state is retained across
invocations as long as the container is reused, but is lost when the
FaaS platform shuts down a container.
Since the platform is free to shut down a container at any time,
functions must not rely on the persistence of global state for
correctness.

These properties afford FaaS providers a high degree of flexibility in
placing, scheduling, and dynamically replicating function activations.
In the context of \sys, these properties imply that each reused
container has well-defined points in its life cycle---namely between
sequential activations---when its state can be safely restored to a
point after the initialization of global state but before the first
function activation, thereby ensuring sequential request isolation.

\parab{Small write sets} FaaS functions, particularly those written in
managed languages, often use a substantial amount of memory, but only
a small proportion of it is modified during an activation. \update{This improves \sys's efficiency because only modified parts of memory need to be restored after an activation.} Our
empirical evaluation on 58 benchmarks shows that the number of pages
actually modified by each function invocation is only a small fraction
of the overall function memory (mean: 8.5\% of the mapped address
space is modified, median: 3.3\%, 90p: 17\%). \resolvepeter{A similar
  observation was reported by REAP \cite{REAP} where the examined functions'
  working sets \finalpass{(both only read as well as modified pages)} were on average 9\%
  of their memory footprints.}

\subsection{Design options}
\label{sec:design-options}

Besides the trivial solution of using a fresh container for every
request, which is inefficient, there are three broad design
approaches for efficient sequential request isolation.

\parab{Language-based approaches} When using appropriate safe
programming languages~\cite{boucher2018putting}, compiler
instrumentation techniques~\cite{vogt2015lightweight}, or
runtimes~\cite{wasm} to implement functions, the language semantics
can ensure efficient request isolation. However, this approach
requires all tenants to use a particular (set of) programming
languages/compilers, prevents the use of libraries written in unsafe
languages for efficiency, and is vulnerable to bugs in the language
runtime.

\parab{Fork} A simple process-based technique is to {\tt fork} a fully
initialized function process, execute an activation within the child
process, and discard the child process after the activation
finishes. The main limitation of this approach is that {\tt fork} as
implemented in general-purpose operating systems cannot capture the
state of a multi-threaded process. To take full advantage of container
reuse, we need to be able to snapshot the fully initialized runtime of
a managed language like  JavaScript, which typically includes
\finalpass{multiple} active threads. \finalpass{Additionally, {\tt fork} (or any copy-on-write (CoW) based approach) incurs expensive data-copying page faults during the execution of the function (i.e., on the critical path of a request).}

\parab{Custom snapshot/restore facilities} have been explored in prior
work
~\cite{faasnap,catalyzer,REAP,Stenbom2019refunction,prebaking,pagurus}
to reduce container cold-start costs by snapshotting an initialized runtime
to disk/memory, and restoring it when a new container is needed. In
principle, this approach could be used to instantiate a container for
each activation. While substantially better than a cold-start for each
activation, instantiating a container from a snapshot is still too
expensive when compared to container reuse for many functions in our
benchmarks.

\subsection{Threat model}
\label{sec:threat}

The FaaS platform, including the platform software, OS kernels,
hypervisors, and platform services are trusted. We assume that the
platform authenticates clients who connect to HTTP/S endpoints, and
enforces access control to functions, as well as a function
activation's access to platform services according to the
authenticated client's credentials.

Legitimate tenants are expected to set up access control lists that
allow only legitimate parties to invoke their functions, and prevent
unwanted information flow via platform services among legitimate
callers with different privileges.

Function implementations provided by tenants, including any libraries
they link and the language runtimes they rely on, are untrusted and
may contain bugs.  Under these assumptions, {\sys} \update{prevents
  leaks of information from a function activation to subsequent ones,
  while} allowing container reuse.

\section{\sys Design}

\begin{figure}[h]
	\centering
	\includegraphics[clip, trim=2.4cm 3.6cm 10cm 9.1cm,width={.45\textwidth}]{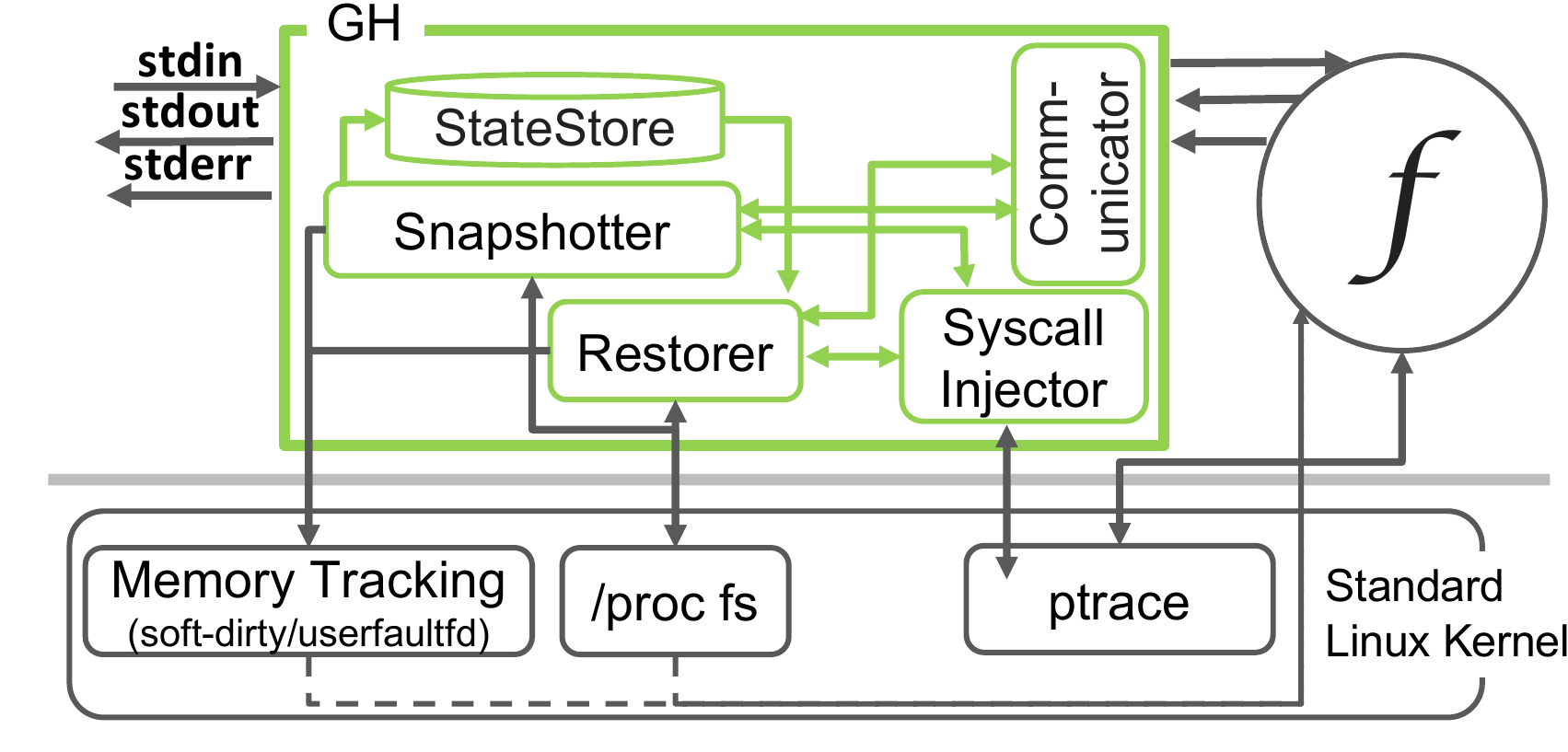}
	\figcap{\sys Architecture: (1) The \manager (in green), (2) The function process. %
		\sys relies on standard Linux kernel utilities.}
	\label{fig:arch}
\end{figure}

\autoref{fig:arch} illustrates the \sys architecture.  The \sys
\emph{manager} process (outermost green box) runs within an OS container %
alongside the function process, and is responsible for enforcing
request isolation.
\sys uses a novel, light-weight in-memory process snapshot/restore
facility that achieves low restore times. The facility relies on
standard Linux kernel facilities to snapshot, track, and restore
processes. The design was guided by the following goals:

\parab{Generality} The facility operates on a generic, multi-threaded
POSIX process and does not make any assumptions about the function
executing inside the process.

\parab{Restore cost proportional to modified pages} To take
advantage of the fact that most function activations only modify a
small proportion of the function process' state, \sys tracks which pages are modified
during a function activation using the Linux soft dirty-bit
facility. As a result, \sys need only restore modified pages after an
activation.

\parab{Restore cost off the critical function execution path} FaaS
platform servers are less than fully utilized most of the
time. Therefore, the design of \sys's snapshot/restore facility seeks
to minimize overhead \emph{during} a function's execution, in favor of
performing all restore-related tasks \emph{between} function
activations. In particular, \sys avoids copy-on-write, which
would burden function execution with expensive data-copying page
faults. As a result, \sys avoids overhead on the end-to-end function
execution latency in the common case of a less than fully utilized
server.

\subsection{Container initialization}

\resolvepeter{The \sys manager process interposes between the Faas
  platform and the process executing the function. The FaaS platform
  initializes the \sys manager process as if it were the process
  executing the function.  The \sys manager then receives requests
  from the FaaS platform, relays them to the function process, and
  communicates results back to the FaaS platform. It communicates with
  the FaaS platform using the latter's standard communication
  mechanisms (in \ow -- the platform on which our prototype runs --
  these communication mechanisms are usually stdin and stdout pipes.)}

\update{
To initialize the actual function process, \sys forks a new process,
prepares pipes for communicating with it, drops privileges of the
child process, and execs the actual function runtime in the child
process.

Next, \sys creates a snapshot of the function process. For this, \sys
invokes the function with \emph{dummy} arguments that are independent
of any client secrets \finalpass{(the dummy arguments are provided by the
  function deployer, once for every function they deploy, and can be
  part of the function's configuration)}. After the function returns,
it snapshots the state of the function process as described in
\autoref{sec:design:snapshotting}. The purpose of the dummy invocation
is to trigger lazy paging, lazy class loading and any
application-level initialization of global state, and to capture these
in the snapshot. Snapshotting without a dummy request would cause
these (expensive) operations to happen again after every state
restoration, which would increase the latency of subsequent function
activations. This is particularly relevant when the function runs in
an interpreted runtime like Python or \nodejs, which heavily rely on
lazy loading of classes and libraries~\cite{lion2016don}.

After this snapshot is created, {\sys} informs the platform that
it is ready to receive actual function invocation requests.
}

\subsection{Snapshotting the function process}
\label{sec:design:snapshotting}

To take a snapshot, the manager interrupts the function process, then:
(a) stores the CPU state of all threads using ptrace~\cite{ptrace};
(b) scans the \texttt{/proc} file system to collect the memory mapped
regions, memory metadata, and the data of all mapped memory pages; (c)
stores all of this in the memory of the manager process; and (d)
resets the soft dirty bit memory tracking state. \finalpass{ Finally, the manager resumes the function process, which then waits for the first request inputs. Subsequently, \sys would restore the function's process state back to this snapshot before each new request.}

\subsection{Tracking state modifications}
\label{sec:design:tracking}

\sys uses the standard \emph{Soft-Dirty Bits} (SD) feature of the 
Linux kernel~\cite{sd-bits}\footnote{\resolvepeter{Available on stock Linux kernels v3.11+. We identified and reported a bug that affected the accuracy of the SD-Bits memory tracking in v5.6, it was fixed in
v5.12~\cite{soft-dirty-bug}}.}, which provides a page-granular, lightweight
approach to tracking memory modifications.  Each page has an
associated bit (in the kernel), initially set to 0, that is set to 1
if the page is modified (dirtied).  When a function invocation
completes, \sys scans the SD-bits exposed by the Linux /proc
filesystem to identify the modified pages.  After restoring the
function process, \sys resets all SD-bits to 0, ready for the next
invocation.

We considered using Linux's user-space fault tracking file descriptor
(UFFD)~\cite{uffd}\footnote{\resolvepeter{Write protection notifications available on stock Linux kernels v5.7+.}} feature for memory tracking and prototyped this
alternative; however we found UFFD to have significantly higher
overhead compared to SD-bits due to the frequent context switches to
user-space for fault handling.\footnote{A custom in-kernel facility
  that allows an application to request a list of modified pages
  presumably could be much faster, but would require kernel
  modifications.} UFFD was faster than soft-dirty bits (and that too
marginally) only when the number of dirtied pages was close to zero.

\subsection{Restoring to the snapshotted state}
\label{subsec:restoration}

When a function invocation completes, the function process returns the
result to the \sys manager.  \sys's manager awaits the function
response and forwards it to the FaaS platform (which then sends it to
the caller).  Next, the manager interrupts the function process and
begins a restore.  The manager identifies all changes to the memory
layout by consulting \texttt{/proc/pid/maps} and \texttt{pagemap}
(e.g. grown, shrunk, merged, split, deleted, new memory regions);
these changes are later reversed by injecting syscalls using
ptrace~\cite{ptrace} (described below).  The manager restores
\texttt{brk}, removes added memory regions, remaps removed memory
regions, zeroes the stack, restores memory contents of pages that have
their SD-bit set, restores registers of all threads, madvises newly
paged pages, and finally resets SD-bits.

To reverse memory layout changes, \sys injects system calls in the
function process using
ptrace~\cite{ptrace,ptrace_do,Stenbom2019refunction}.
After restoration completes, the child process is in an identical
state to when it was snapshotted, and the process is ready to execute
the next request.  \update{As an optimization, if a function within
  the same container is invoked consecutively by mutually trusting
  callers, then \sys\ can skip the rollback between such invocations.}

\subsection{Enforcing request isolation}
\label{sec:design:security}
\resolvepeter{\sys enforces request isolation by design. \sys prevents
  new requests from reaching the function's process until it has been
  restored to a state free from any data of previous requests. This is
  achieved by intercepting the end-client requests before they reach the
  function and buffering them in \sys until the function's process
  has been restored.}

\resolvepeter{Although intercepting the communication ensures control
  of the function process and \finalpass{enforces} security, it can add an overhead
  of copying request input/outputs to and from \sys's manager
  process. This overhead can be eliminated as follows: (1) The FaaS
  platform can forward inputs directly to the function process after
  waiting for a signal from \sys's manager process that the function
  has been restored to a clean state. This requires minor, but trusted
  changes to the FaaS platform to wait for the signal from \sys. (2)
  Upon completion of a request, the function process can return
  outputs directly to the FaaS platform and, separately, signal \sys's
  manager process that its state can be rolled back. The changes
  needed can be made in the I/O library that handles communication with the platform in the function process (no
  changes needed to the code of the individual functions submitted by the developers) and the changes
  do \emph{not} have to be trusted for security -- if the function
  process fails to signal \sys that it is done, \sys will never
  restore the state and will never signal the platform to send the
  next request.\footnote{We implemented (2) to facilitate debugging. Our evaluation still intercepts all inputs and outputs to demonstrate that platform modifications were not required and show the overhead of such interception on various functions.}}

\parab{Design limitations:} \sys does not currently allow function
implementations to open arbitrary network connections and file
descriptors.  (None of the benchmarks we use in the evaluation require
them.)  Instead, functions are expected to rely on platform services
for network communication and storage.
\update{ Additionally, functions must externalize any caches whose
  state they wish to retain across requests since the state of
  in-address-space caches will be rolled back at every restore.}
\update{Finally, as stated in our threat model, any external
  state (e.g. external storage, or the state of network connections
  and pipe contents) is assumed to be subject to access control. This
  is necessary to prevent data leaks across clients with different
  privileges via the external state.}

\newcommand{\lsys}{\textsc{gh}\xspace}
\newcommand{\Lsys}{\textsc{gh}\xspace}
\newcommand{\lsysnop}{\textsc{gh$_{\hspace{-0.5mm}\text{\relsize{-1}{\textls[-100]{nop}}}}$\xspace}}
\newcommand{\lfork}{\textsc{fork}\xspace}
\newcommand{\Lfork}{\textsc{fork}\xspace}
\newcommand{\lbase}{\textsc{base}\xspace}
\newcommand{\lfaasm}{\textsc{faasm}\xspace}

\definecolor{lsyscolor}{HTML}{5e3c99}
\definecolor{lsysnopcolor}{HTML}{b2abd2}
\definecolor{lforkcolor}{HTML}{ffd096}
\definecolor{lfaasmcolor}{HTML}{e66101}

\newcommand{\lsysshaded}{%
\raisebox{-1mm}{\begin{tikzpicture}[
  x=\textwidth/100,y=\textwidth/100,
  every node/.style={inner ysep=0.2,outer sep=0, inner xsep=1},
  ]
\node [minimum height=4mm, fill=lsyscolor, draw=none, text=white] at (0,0) {\lsys};
\end{tikzpicture}}\xspace}

\newcommand{\lsysnopshaded}{%
\raisebox{-1mm}{\begin{tikzpicture}[
  x=\textwidth/100,y=\textwidth/100,
  every node/.style={inner ysep=0.2,outer sep=0, inner xsep=1},
  ]
\node [minimum height=4mm, fill=lsysnopcolor, draw=none] at (0,0) {\lsysnop};
\end{tikzpicture}}\xspace}

\newcommand{\lforkshaded}{%
\raisebox{-1mm}{\begin{tikzpicture}[
  x=\textwidth/100,y=\textwidth/100,
  every node/.style={inner ysep=0.2,outer sep=0, inner xsep=1},
  ]
\node [minimum height=4mm, fill=lforkcolor, draw=none] at (0,0) {\lfork};
\end{tikzpicture}}\xspace}

\newcommand{\lfaasmshaded}{%
\raisebox{-1mm}{\begin{tikzpicture}[
  x=\textwidth/100,y=\textwidth/100,
  every node/.style={inner ysep=0.2,outer sep=0, inner xsep=1},
  ]
\node [minimum height=4mm, fill=lfaasmcolor, draw=none] at (0,0) {\lfaasm};
\end{tikzpicture}}\xspace}

\section{Evaluation}
\label{sec:evaluation}

In this section we evaluate {\sys}'s performance across a wide range
of FaaS benchmarks.  Overall, we show that:
\begin{itemize}
\item For a wide range of benchmark functions using three different
  languages/runtimes, {\sys} has modest overhead on end-to-end
  latency and throughput.
\item {\sys}'s latency overhead depends primarily on the memory
  characteristics of the function and is proportional to the number of
  pages dirtied during a function's execution. {\sys}'s throughput
  scales nearly linearly with the number of available cores.
\item {\sys}'s lightweight restoration has equivalent or better
  performance than an alternative fork-based isolation approach, which
  is less general. We also compare to a \wasm-based isolation
  approach, and show that \sys\ has competitive performance despite
  being more general.
\end{itemize}

\subsection{Evaluation Overview}

\parab{Implementation.}  We implemented \sys in \textasciitilde7K
lines of C.  {\sys} is compatible with off-the-shelf Linux and
requires no kernel changes.%

\parab{OpenWhisk Integration.}
\label{eval:integration} 
We integrated \sys with \ow~\cite{ow-commit}, by modifying \ow's container runtimes for Python and {\nodejs} to include \sys.  In addition, we implemented an \ow container runtime for native C, to enable the evaluation of
native C FaaS benchmarks. \resolvepeter{Most \ow runtimes use the actionloop-proxy design, where a distinct process acts as a proxy that communicates with the \ow platform (through HTTP connections), and forwards the requests to the runtime process (through stdin) which has a simple wrapper to process inputs, call the developer's function, and return results. \sys interposes between the proxy and the runtime, intercepting the stdin and stdout and forwards the stdin only when the function's process is restored to a clean state. \ow's container runtime for \nodejs, on the other hand, is built using a single process that directly interacts with the platform and runs the function. We refactored it to an actionloop-proxy-like design to maintain a uniform \sys implementation that ensures security by blocking inputs until the function's process is restored to a clean state\footnote{Encapsulating the full process would require \sys to implement the platform API or have a small platform modification to allow blocking inputs until \sys signals to the platform that the function's process is being restored as described in \ref{sec:design:security}.}}.

\parab{Hardware Configuration.} We run all experiments on a private
cluster hosting OpenStack/Microstack (ussuri, revision 233). \update{Each
physical host has an Intel Xeon E5-2667 v2 2-socket, 8-cores/socket
processor, 256GB RAM and a 1 TB HDD.}

\parab{OpenWhisk Deployment.}  \update{We use the standard distributed
  Openwhisk deployment. Our distributed setup comprises 2VMs. One VM
  runs all \ow core components except for the invoker, which runs on
  a separate VM. The invoker is the component responsible for starting
  function containers locally and dispatching function requests to
  them; this is the component that interacts with and hosts \sys. We
  choose to isolate the invoker component in a separate VM to have
  more control on the variables affecting the experiments.

Both VMs are placed on the same physical host to minimize network
communication overhead, creating favorable baseline conditions. To
reduce potential performance interference, we pin the two VMs to
separate cores and ensure that their memory is allocated from the
corresponding NUMA domain. VMs are configured with 64GB RAM and an
experiment-dependent number of cores (SMT turned off). The VMs run
Ubuntu 20.04 with a stock Linux kernel v5.4. \ow is configured to 
run all functions with a 2GB RAM limit and a 5 minute timeout.}

\parab{Experiment Configurations.}  To evaluate {\sys}'s overheads, we
run two primary configurations: \textbf{\lbase}, an insecure baseline
using unmodified \ow that does not provide sequential request isolation \update{(we prevent container cold-starts in our experiments to deliberately create an unfavorable but conservative baseline)};
and \textbf{\lsys}, which uses {\sys} on OpenWhisk to provide
sequential isolation.

\update{We also run a third configuration \textbf{\lsysnop}, which
  includes {\sys} but does not restore dirtied pages between
  consecutive invocations of the same function. This configuration
  represents an optimization for the case where consecutive requests
  are from the same security domain. The configuration also helps delineate
  {\sys}'s page tracking and restoration %
  costs, which is the difference between the {\lsys}
  and {\lsysnop} configurations.}

Lastly, we compare {\sys} to two alternative approaches.
In \autoref{sec:eval:fork} we implement a fork-based request isolation
method, \textbf{\lfork}, which is applicable to single-threaded
applications and runtimes only.  In~\autoref{sec:benchmarks:faasm} we
compare {\sys} to \textbf{\lfaasm}, a research FaaS platform designed
to reduce cold-start latencies for {\Wasm}-compatible functions.  We
detail these alternative approaches in the respective sections.

\subsection{Microbenchmarks}
\label{subsec:microbenchmarks}
\begin{figure}[tb]
  \centering
    \includegraphics[clip, trim=0.3cm 0.3cm 0.4cm 0.3cm,width={0.25\textwidth}]{{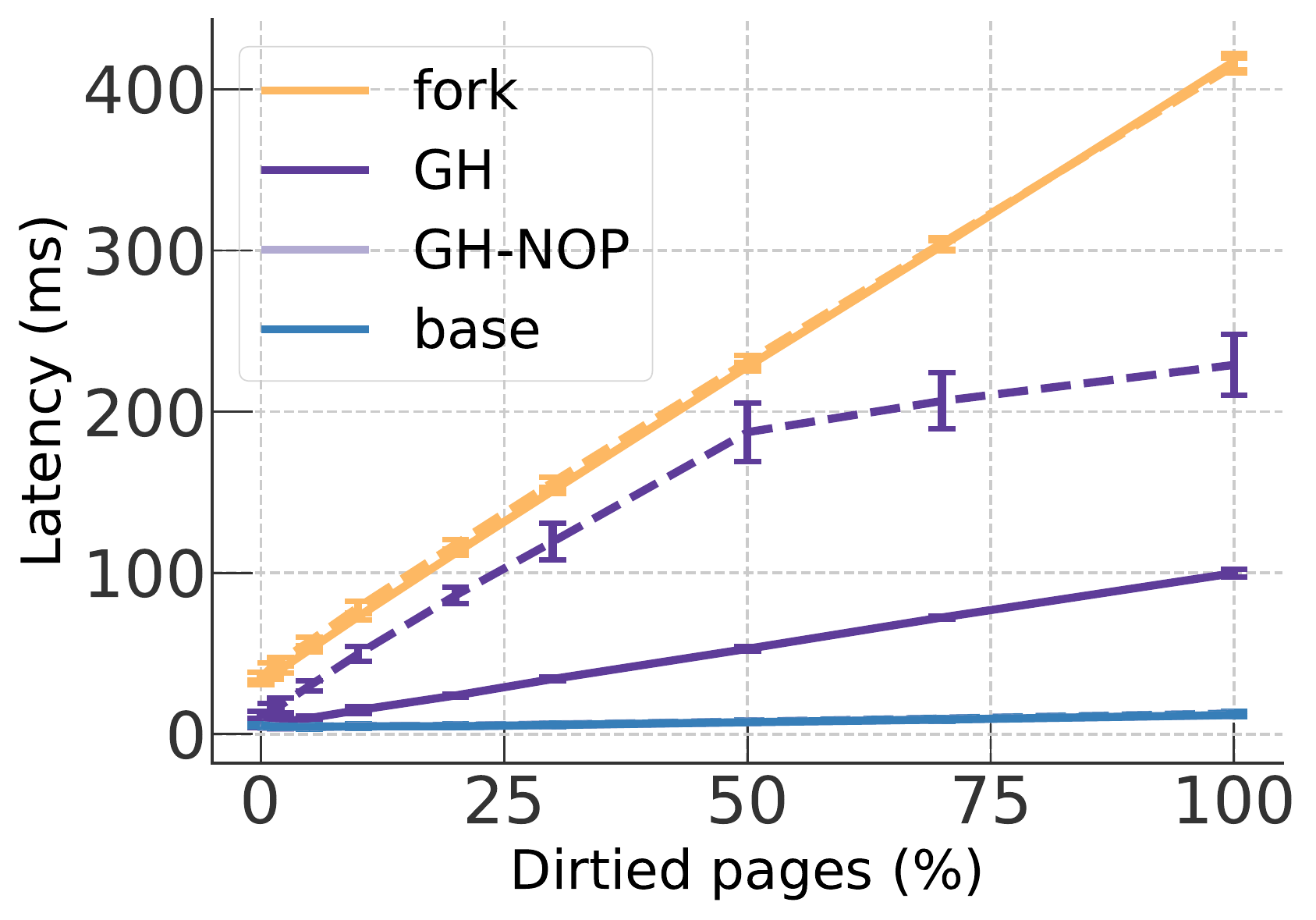}}
    \includegraphics[clip, trim=1.2cm 0.3cm 0.4cm 0.3cm,width={0.22\textwidth}]{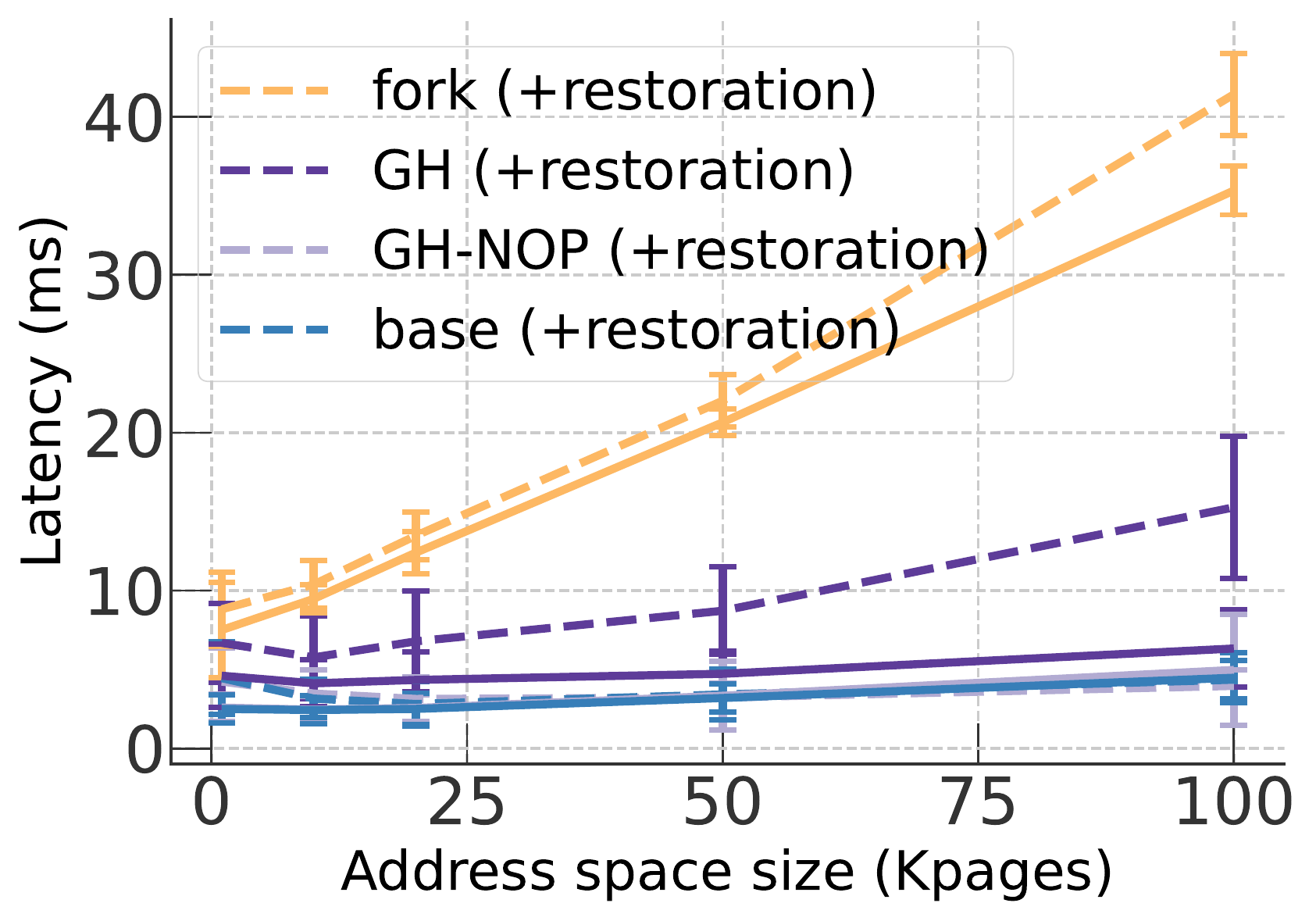}
  \caption{Latencies varying the number of dirtied pages (left) and
    the address space size (right). Different colors represent
    different request isolation methods (or no isolation for
    {\lbase}). Solid lines are latencies with in-function overhead but
    not restoration overhead, while dashed lines include both. (The lines of {\lbase} and
    {\lsysnop} coincide visually in the figure.)}%
  \label{fig:microbenchmarks-invoker}
  \label{fig:micro-vary-dirty}
  \label{fig:micro-vary-vm-size}
\end{figure}

In this experiment, we evaluate {\sys}'s impact on request latency and
how that impact varies with the memory size and the number of pages
dirtied\footnote{We also considered address space fragmentation (same
overall address space size but a varying number of memory maps) as an
independent variable, but found that it has no statistically
significant impact on the overhead of \lsys or \lfork.}.  We evaluate
both the \emph{in-function} overheads that are on the critical path of
function execution, and the \emph{restoration overhead} which occurs
off the critical path.  We defer evaluating
{\sys}'s snapshotting overheads, which occur only once after a new
container starts, to~\autoref{subsec:snapshotting}.

\parab{Microbenchmark.} We implement a simple function in C that
pre-allocates an address space of a fixed size.  Each invocation
(a) dirties a subset of the pages by writing a word to each page of
that subset, then (b) reads one word from each mapped page, even those that were not dirtied.  We set up a
4-core VM with a single function-hosting container (this container
is limited to 1 core), initialize the container, and then repeatedly
invoke the function. We measure function latencies at the \ow invoker.

\subsubsection{In-Function Overheads}~\\
\label{sec:eval:infunctionoverheads}
\parab{Low-load Workload.} We run the microbenchmark with a \emph{low
load} workload comprising 150 requests submitted one-at-a-time, with a
small delay between consecutive requests.  This delay is sufficient
for {\sys} to complete restoration before the next request arrives, so
measurements for the low-load workload capture only the in-function
overheads.

\parab{Results.} The solid lines in \autoref{fig:micro-vary-dirty}
(left) plot function latency as we vary the percentage of pages
dirtied from 0\% to 100\% with a fixed 100K mapped pages.  As
expected, \lsys introduces some latency overhead proportional to the
number of dirtied pages.  This overhead is due to a minor page fault
to set the soft-dirty (SD) bit when a page is dirtied, which is required by the
SD-bit mechanism on our hardware.  In contrast, {\lsysnop} has
negligible overhead relative to \lbase since the SD-bits set in the first
run are not reset (there is no memory restoration), and thus these
page faults are not incurred in subsequent runs.

We also run a variant of the experiment where we fix the number of dirtied pages to 1K and vary the address space size from 1K to 100K pages.  The solid lines in \autoref{fig:micro-vary-dirty} (right) show the function latency as we vary the address space size.  We observe that \lsys's overhead is constant with respect to address space size because the in-function overheads depend only on the number of dirtied pages, which is fixed now.

\subsubsection{In-function + Restoration Overheads}~\\
\label{sec:eval:restorationoverheads}
\parab{High-Load Workload.}  We repeat the two experiments above with
a \emph{high load} workload comprising 150 requests submitted
back-to-back with no delay between consecutive requests.  This leads
to additional delays while waiting for {\sys} to complete restoration
after the previous request.  In contrast to the low-load workload, the
high-load workload thus reflects \emph{both} the in-function
\resolvepeter{and the off-critical-path} restoration overheads.

\parab{Results.}  The dashed lines in \autoref{fig:micro-vary-dirty} (left) show the function latency as we vary the percentage of pages dirtied from 0\% to 100\% with a fixed 100K mapped pages.  We observe higher latency overheads for the high-load workload (dashed lines) compared to the low-load workload (solid lines), and these overheads grow linearly as the percentage of dirtied pages increases.  There is a change in slope at 60\% because \sys is able to coalesce individual page restorations into fewer, larger memory copy operations, which are more efficient.

Next, we repeat the second experiment variant.  \autoref{fig:micro-vary-dirty} (right) shows the function latency as we vary the address space size from 1K to 100K pages while fixing the number of dirtied pages to 1K.  Although in-function overheads are constant, restoration overheads in this experiment increase linearly with the address space size, because during restoration GH must scan the SD-bits of the whole address space to determine the pages to restore.

\subsubsection{Comparison to Fork}
\label{sec:eval:fork}

A potential alternative to our lightweight restoration is to use
copy-on-write techniques such as fork (\autoref{sec:design-options}).
Fork is not general purpose -- it only works for single-threaded
functions -- however we provide a performance comparison for the
purpose of illustration.  We implement fork-based isolation and repeat
the two microbenchmark experiments.  \resolvepeter{In our fork-based
  implementation, we initialize the function up to the same point
  where {\sys} takes its snapshot (a safe clean state).  Instead of
  lightweight restoration, each request is then handled by a separate
  copy of the process forked at that state.}

\autoref{fig:micro-vary-dirty} (left) shows the function latency of \lfork as we vary the percentage of pages dirtied from 0\% to 100\% with a fixed 100K mapped pages.  We observe that the overhead of \lfork is higher than \lsys because each page fault is significantly more expensive than for \lsys, entailing an additional page copy.

\autoref{fig:micro-vary-dirty} (right) shows the function latency of
\lfork as we vary the address space size from 1K to 100K pages while
fixing the number of dirtied pages to 1K.  \update{We see
  significantly higher overhead for \lfork compared to \sys, and a
  linear increase in latency with the address space size. This increase is predominantly due to the additional overhead caused by dTLB misses on the first accesses to each page (even if unmodified) of the new process. This access can additionally require lazy creation of physical page table entries depending on the memory layout of the program.}

\subsection{FaaS Benchmarks}

In this section we evaluate {\sys}'s impact on request latency and
throughput for a range of FaaS benchmarks written in three different
languages.  We first compare {\sys} to an insecure baseline in
OpenWhisk (\autoref{sec:benchmarks:baseline}).  We then provide an
illustrative comparison to a fork-based implementation
(\autoref{sec:benchmarks:fork}) and to \faasm, an alternative
\wasm-based FaaS platform designed to optimize cold-starts, but that
can also be used for request isolation in limited cases
(\autoref{sec:benchmarks:faasm}).

\parab{Benchmarks.} We evaluate 58 functions across three benchmarks and three languages: 22 python functions from the pyperformance benchmark~\cite{pyperf},  23 C functions from PolyBench~\cite{polybench}, and 13 functions (6 python, 7 {\nodejs}) from the FaaSProfiler benchmark suite~\cite{faasProfiler}.

\update{These functions cover a wide variety of real FaaS use cases
  such as Web applications, JSON and HTML parsing/conversion, string
  encoding, data compression, image processing (2D, 3D), optical
  character recognition (OCR), sentiment analysis, matrix computations
  (e.g. multiplication, triangular solvers), and statistical
  computations.}

\parab{Latency.}  To measure latency we deploy a 4-core VM with a
single function container that is limited to at most one core, and run
a closed-loop client in a separate VM on the same machine\footnote{\resolvepeter{This placement minimizes network latencies to achieve best baseline performance and to allow easy and efficient scheduling of our 530 configurations on our resources.}},%
 which submits requests one-at-a-time.  This
workload is similar to the \emph{low-load} setting
from~\autoref{sec:eval:infunctionoverheads} and enables {\sys} to
complete restoration in between consecutive requests, so latency
measurements reflect {\sys}'s in-function overheads only. We report
two latency measurements: the end-to-end latency of requests as
experienced by the end-client (including all FaaS platform delays);
and the invoker latency, which measures only the function execution
time at the invoker, excluding overheads of the remaining FaaS
platform \finalpass{components}, which {\sys} does not affect at all.  All measurements are
averages of 1,200 invocations, except for C functions longer than 10
seconds, where we report averages of 90 invocations.

\parab{Measuring Throughput.} To measure throughput we deploy a 4-core
VM with 4 function containers %
in a separate VM that maintains a large number of in-flight requests
(both the number of function containers and in-flight requests are
chosen empirically to maximize throughput).  This workload is similar
to the \emph{high-load} setting
from~\autoref{sec:eval:restorationoverheads} as it ensures the FaaS
platform is always saturated with requests.  Throughput measurements
thus account for {\sys}'s full overheads including both the
in-function overheads and the restoration overheads. \finalpass{Unless otherwise specified, we report the peak sustained xput in 4 runs, each at least 1.5 minutes long.}

\parab{Detailed Measurements.}  
\resolvepeter {In addition to the figures presented in this section,  full measurement data for our benchmarks can be found in \autoref{appendix:grandtables}. Table 1 shows the absolute latency and throughput measurements for the \lbase, \lsys, \lsysnop, \lfork, and \lfaasm configurations. Table 2 shows the relative overheads compared to an insecure baseline. Table 3 shows the relation between the latency, overheads, and throughput of \sys.}

\begin{figure}[htb]%
\centering%
\footnotesize%
\parbox{4mm}{(a)}\raisebox{-0.5\height}{\includegraphics[trim=0 0 0 5, clip, width=0.95\columnwidth]{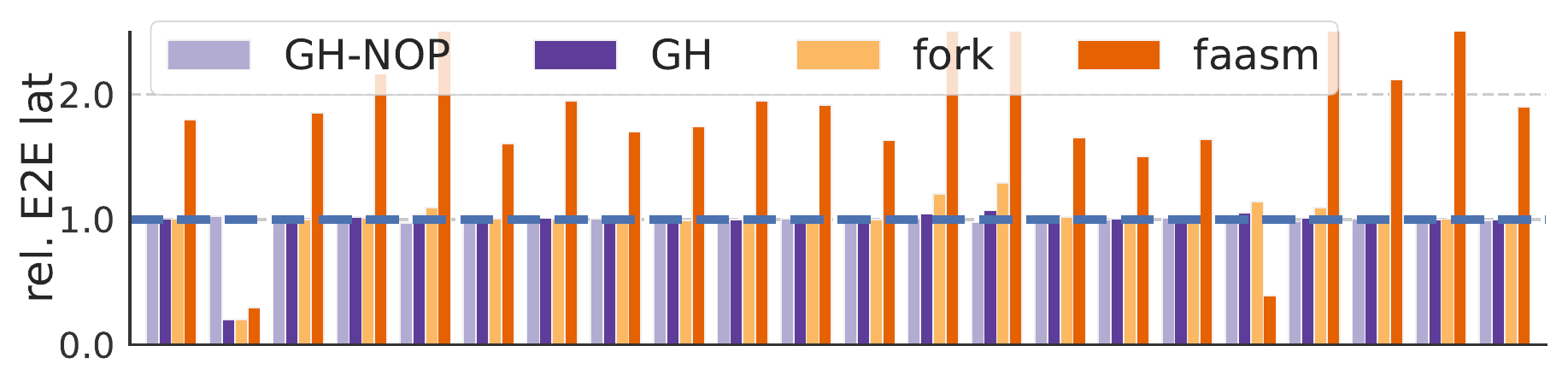}}\\%
\parbox{4mm}{(b)}\raisebox{-0.5\height}{\includegraphics[width=0.95\columnwidth]{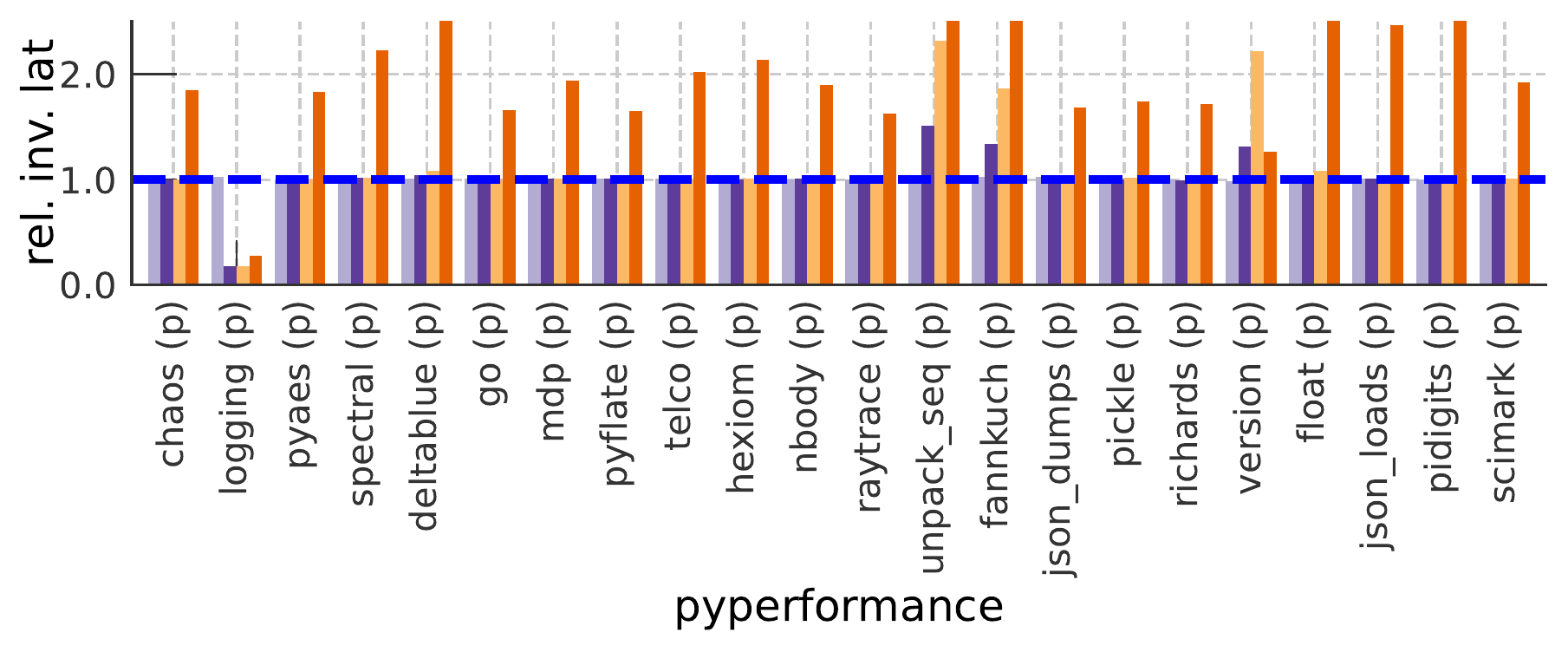}}\\%
\vspace*{-0.1cm}%
\parbox{4mm}{(c)}\raisebox{-0.5\height}{\includegraphics[width=0.95\columnwidth]{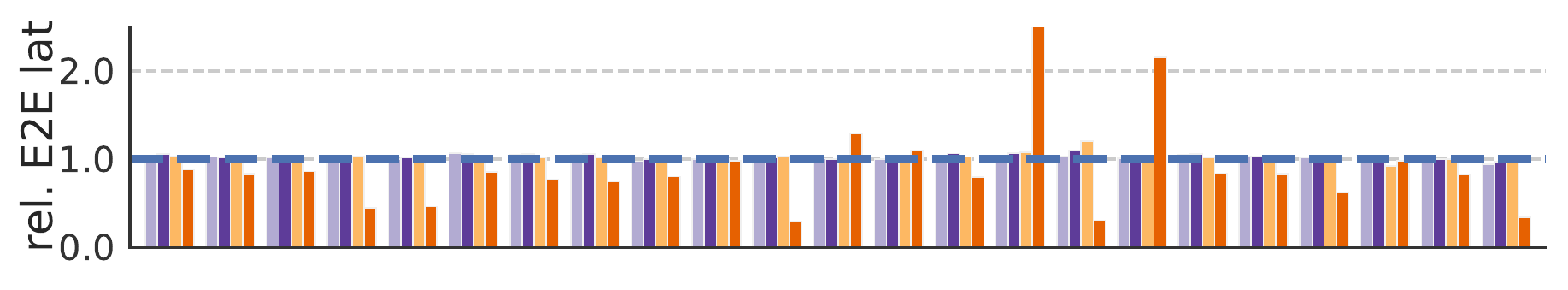}}\\%
\parbox{4mm}{(d)}\raisebox{-0.5\height}{\includegraphics[width=0.95\columnwidth]{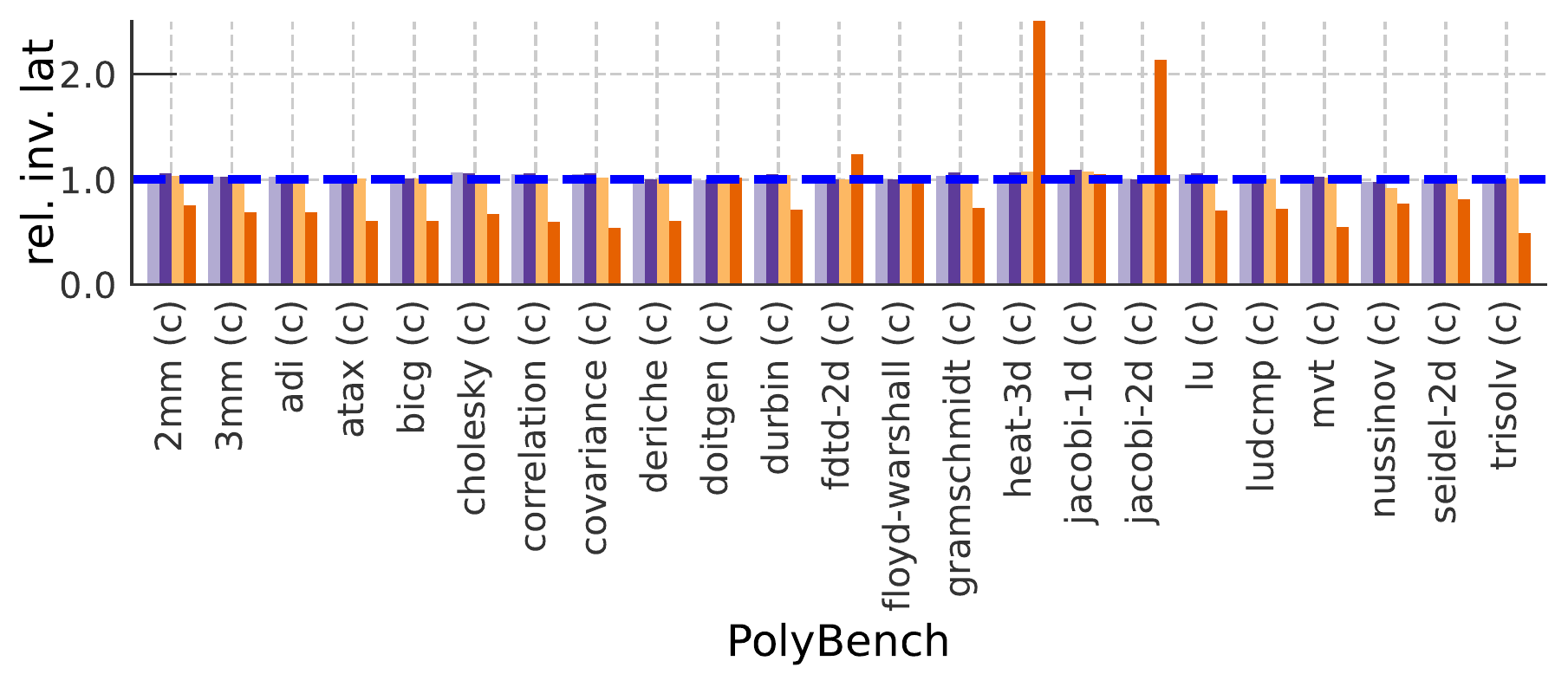}}\\%
\vspace*{-0.1cm}%
\parbox{4mm}{(e)}\raisebox{-0.5\height}{\includegraphics[width=0.4\columnwidth]{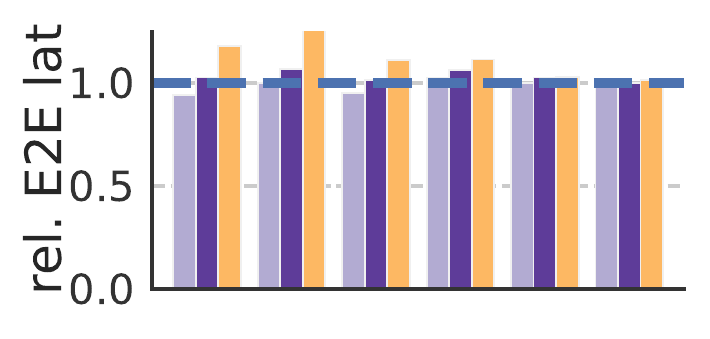}}%
\raisebox{-0.5\height}{\includegraphics[width=0.4\columnwidth]{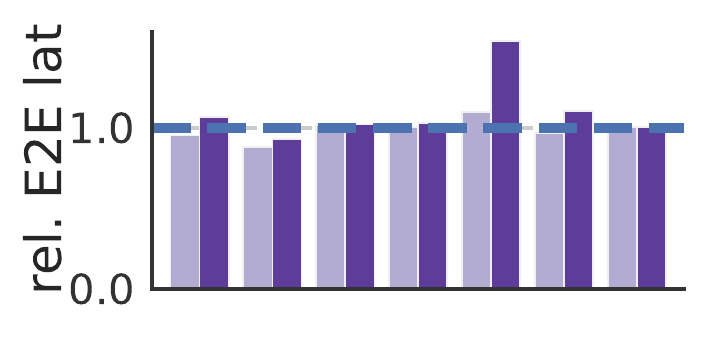}}\\%
\parbox{4mm}{(f)}\raisebox{-0.5\height}{\includegraphics[width=0.4\columnwidth]{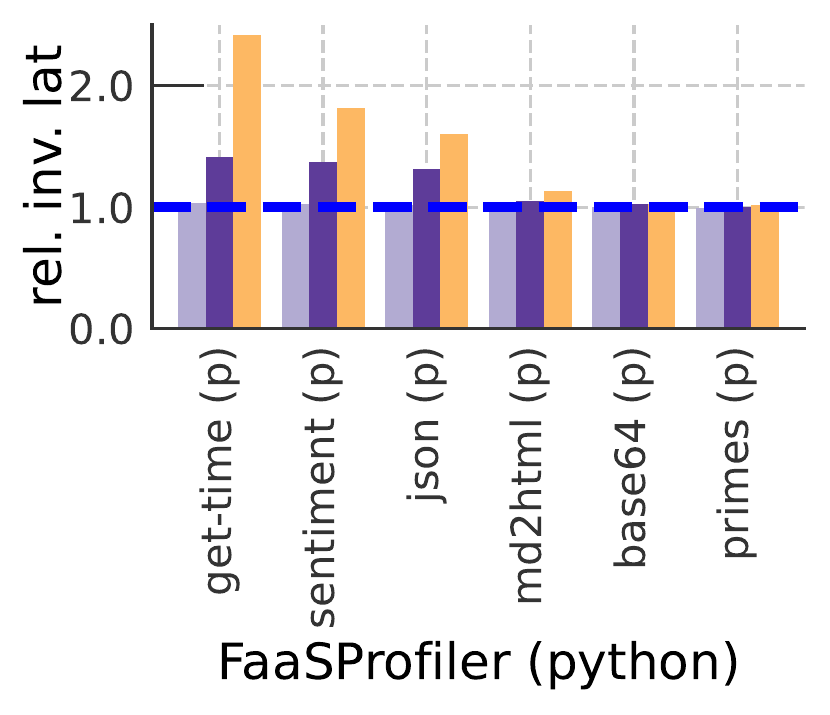}}%
\raisebox{-0.5\height}{\includegraphics[width=0.4\columnwidth]{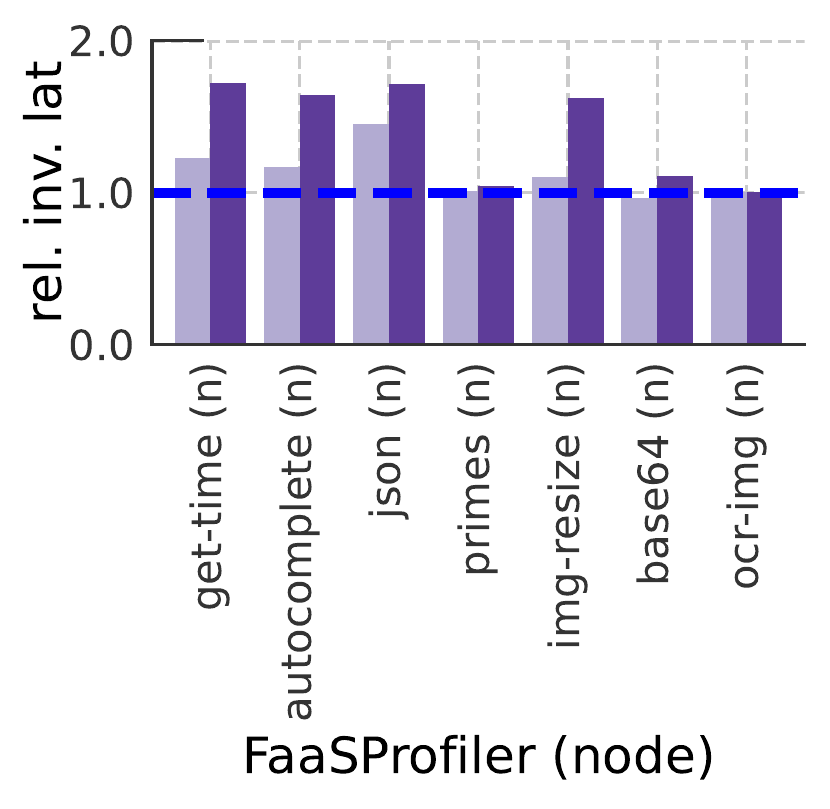}}\\%
\caption{Relative end-to-end latency and invoker-measured latency of \lsys, \lsysnop, \lfork, and \lfaasm compared to the insecure baseline {\lbase}. Figures are capped at 2.5X the baseline. Detailed numbers are in \autoref{appendix:grandtables}.
The symbols (p), (c) and (n) denote Python, C and {\nodejs} benchmarks, respectively. \textbf{Lower numbers are better.}}%
\label{fig:latency}%
\end{figure}

\begin{figure}[htp]%
	\centering%
	\footnotesize%
	\includegraphics[width=0.99\columnwidth]{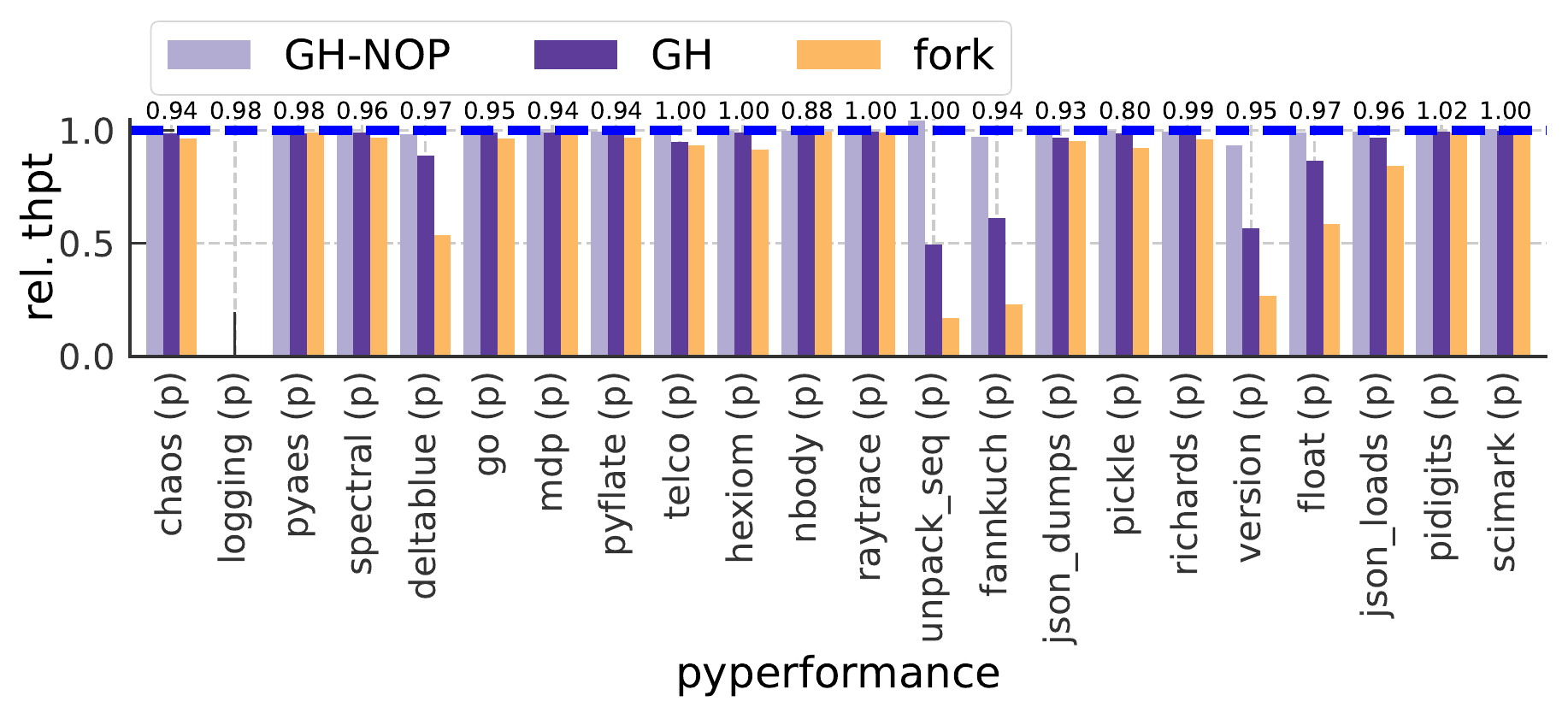}\\%
	\vspace*{-0.1cm}%
	\includegraphics[width=0.99\columnwidth]{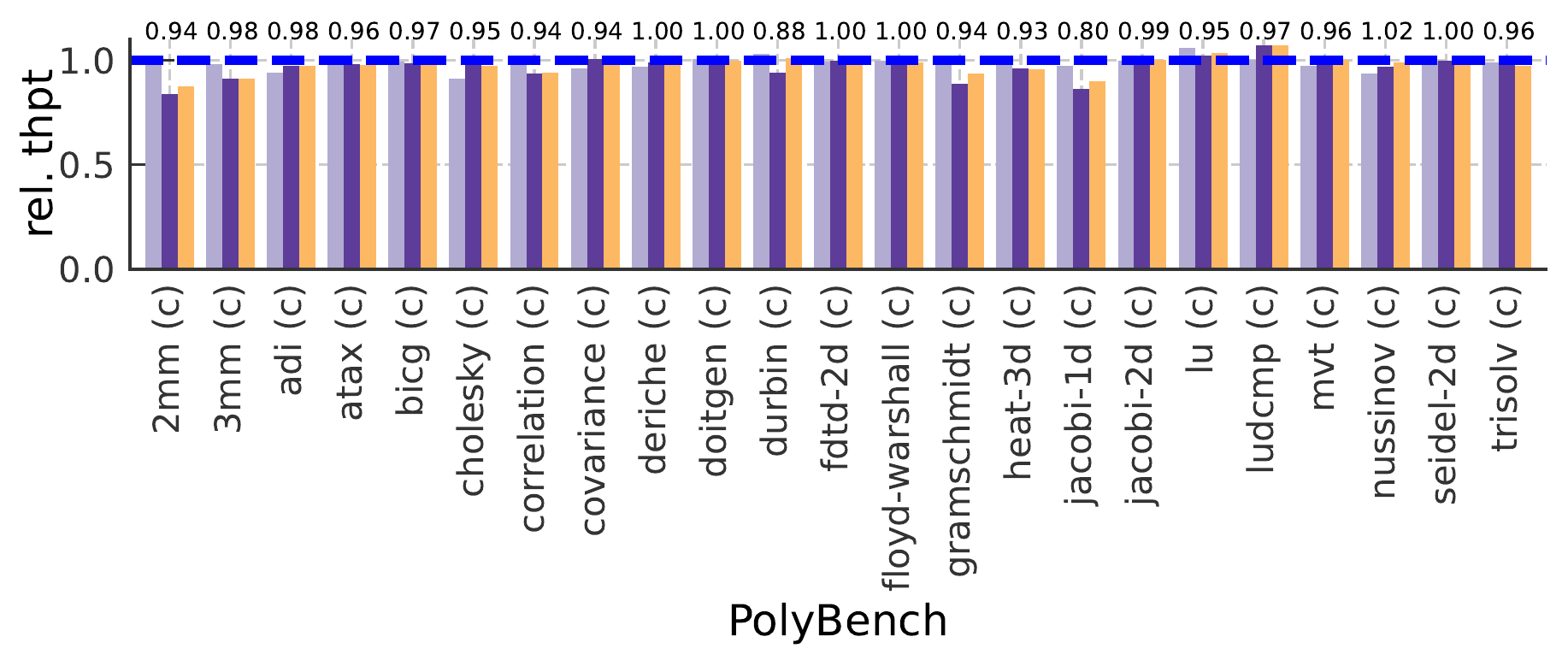}\\%
	\vspace*{-0.1cm}%
	\includegraphics[width=0.45\columnwidth]{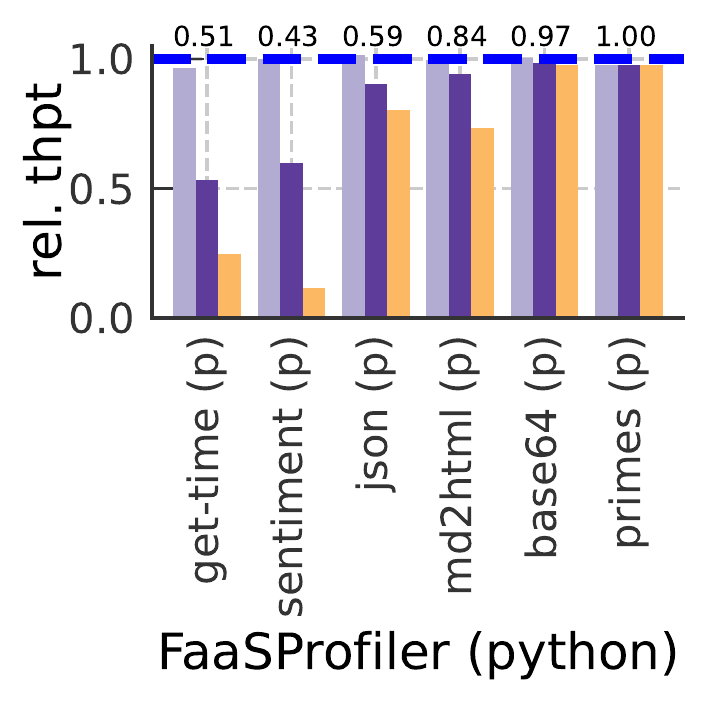}%
	\includegraphics[width=0.45\columnwidth]{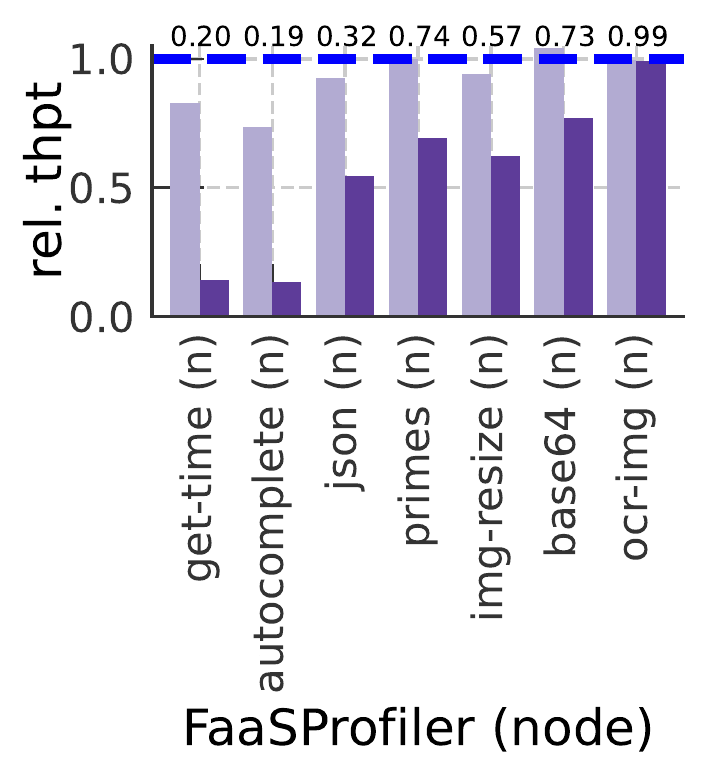}%
	\caption{Relative throughput of \lsys, \lsysnop, \lfork compared to the insecure baseline {\lbase}. Figures are capped at 2.5X the baseline. Detailed numbers are in \autoref{appendix:grandtables}.%
	The symbols (p), (c) and (n) denote Python, C and {\nodejs} benchmarks, respectively. \textbf{Higher numbers are better.}}%
	\label{fig:xput}%
\end{figure}

\subsubsection{Baseline Comparison}~
\label{sec:benchmarks:baseline}
\autoref{fig:latency} (a), (c) and (e) show the end-to-end request latency for all benchmarks.  For each benchmark, we normalize the latency measurements relative to \lbase; thus values \textless 1 indicate better latency than the base line and \textgreater 1 represent worse latency.

We first consider the results for \lsysshaded and \lsysnopshaded.  The
main takeaway is that \lsys overhead on end-to-end latency relative to
\lbase is low overall.  In most cases it is negligible (within one
standard deviation). The median, 95th-percentile and maximum relative
overheads are 1.5\%, 7\% and 54\%, respectively, and the overhead is
below 10.5\% in all benchmarks except img-resize(n), where it is
54.2\% (discussed in the next paragraphs).  %
\update{The low overhead in most benchmarks is unsurprising, because
end-to-end latency measurements include delays within the FaaS
platform that are significant relative to the overhead added by the SD-bit tracking. These significant platform overheads are the same in the baseline and \sys.}

\lsys overheads are more apparent when we inspect invoker latencies.
\autoref{fig:latency} (b), (d) and (f) plot the invocation latency for
all benchmarks, normalized to \lbase.  We observe that for python and
C benchmarks the {\sys} overhead is relatively low.  However, for some
specific {\nodejs} benchmarks (\autoref{fig:latency} (f), right) the
overhead is more pronounced, up to 70\% in the worst case.  This
occurs for two reasons.  

First, \lsys proxies inputs to functions,
which causes additional overheads for some of the {\nodejs} functions
with large inputs such as \texttt{json} and \texttt{img-resize} (which
take inputs of 200kB and 76kB, respectively). \resolvepeter{This cost arises due to our refactoring of \ow's \nodejs runtime wrapper. This overhead can be reduced by integrating \sys with the original single-process version of \ow \nodejs.} %

\resolvepeter{Second, {\nodejs} has a time-dependent behavior in
  garbage collection, namely, garbage collection can be triggered by
  the passage of time. Snapshotting and restoration can adversely
  affect this behavior, because restoration reverts the garbage
  collection state.
  The impact of this garbage collection was particularly pronounced on
  some benchmarks such as img-resize(n). The problem can be alleviated
  by virtualizing time such that the process restoration resets the
  time to the original time of the snapshot, or by modifying the
  garbage collection in a time-independent way. This is actually
  a broader problem in the space of snapshot and restore techniques;
a comprehensive treatment of this topic is beyond the scope of this
paper and left for future work.}

Surprisingly, \lsys is faster than \lbase on the benchmark \texttt{logging(p)}. We
discovered that this occurred due to a memory leak in the function's
original implementation causing it to slow down after repeated
invocations.  \lsys's restoration also rolls back the leaked memory,
thus avoiding the slowdown.

\autoref{fig:xput} shows the request throughput for all benchmarks,
normalized to \lbase.  Since functions are invoked sequentially, the
throughput of \lsys relative to \lbase should be inversely
proportional to \lsys's relative invoker latency, which is
approximately 1 + (in-function overhead + restoration overhead)/(baseline invoker latency). Our observations are consistent with this: The
throughput plots in \autoref{fig:xput} show the reciprocal of this
calculated number above each benchmark, and the heights of the {\lsys}
bars are approximately equal to this reciprocal, as would be
expected. For 40 out of 51 C and Python benchmarks the {\lsys}
throughput is within 10\% of {\lbase}, and up to 50\% lower on the
remaining 11, mostly very short benchmarks. On {\nodejs} benchmarks,
where {\lsys}'s relative invoker latencies can be very high (as
explained above), {\lsys}'s throughput is between 2\% and 86\% less
than {\lbase}'s. \finalpass{{\lsys}'s \nodejs restoration overheads also tend to be higher than other runtimes as \nodejs's runtime maps memory and performs memory layout changes aggressively (see~\autoref{fig:restoration_overheads} for the restoration overheads of specific benchmarks).}  Across all benchmarks, the median and 95th-percentile
throughput reductions are 2.5\% and 49.6\%, respectively.

\subsubsection{Comparison to Fork}
\label{sec:benchmarks:fork}

We also provide a comparison to the \lfork alternative described in~\autoref{sec:eval:fork}.  Recall that \texttt{fork} is only applicable to single-threaded functions, thus we are unable to provide measurements for the {\nodejs} runtime.  

\autoref{fig:latency} also plots results for {\lforkshaded} for single-threaded benchmarks. The latency overhead of \lsys is slightly less than that of \lfork since \lsys's page faults are lighter than those of \lfork (\lfork's page faults also require page copying, while \lsys's page faults only set a SD-bit each).

\autoref{fig:xput} shows that the throughput of \lfork follows a similar rule to that of \lsys.  \resolvepeter{When compared to \lsys, \lfork's throughput is similar on all but very short benchmarks, where \lsys's throughput is noticeably higher than \lfork's.}

\subsubsection{Comparison to Request Isolation using Faasm\\}
\label{sec:benchmarks:faasm}
A potential alternative to \sys's process-based request isolation is to implement request isolation in the language runtime.  To illustrate the performance trade-offs of the two approaches, we
compare \sys to \faasm~\cite{faasm}, a \finalpass{state-of-the art} FaaS platform where functions
are isolated from each other not using OS containers but by compiling
them to {\Wasm}, and relying on spatial isolation within {\Wasm}'s
runtime. \faasm is designed to reduce FaaS cold-start latencies, but
it can be used for efficient request isolation: {\Wasm} limits each
function to a \emph{contiguous} 4GB memory map, which \faasm can quickly
restore simply by a copy-on-write remapping after each request. Note that {\faasm} is not a fully general solution to the request isolation problem since it places
restrictions on the functions -- most notably, they must
compile to {\Wasm}.

\update{{\faasm} comes with its own FaaS
platform, which is significantly different from \ow.
Despite the differences in the platforms, which make a direct
comparison difficult, we compare \sys and \faasm for completeness sake.
For the comparison, we use the pyperformance (python) and PolyBench
(C) benchmarks, both of which can be compiled to {\Wasm} as
demonstrated in \cite{faasm}. We rely on \faasm's microbenchmarking
infrastructure that reports both the overall latency (end-to-end and
invoker) and the restoration (reset) cost.}

\begin{figure}[tb]%
	\centering%
	\footnotesize%
	\includegraphics[width=0.95\columnwidth]{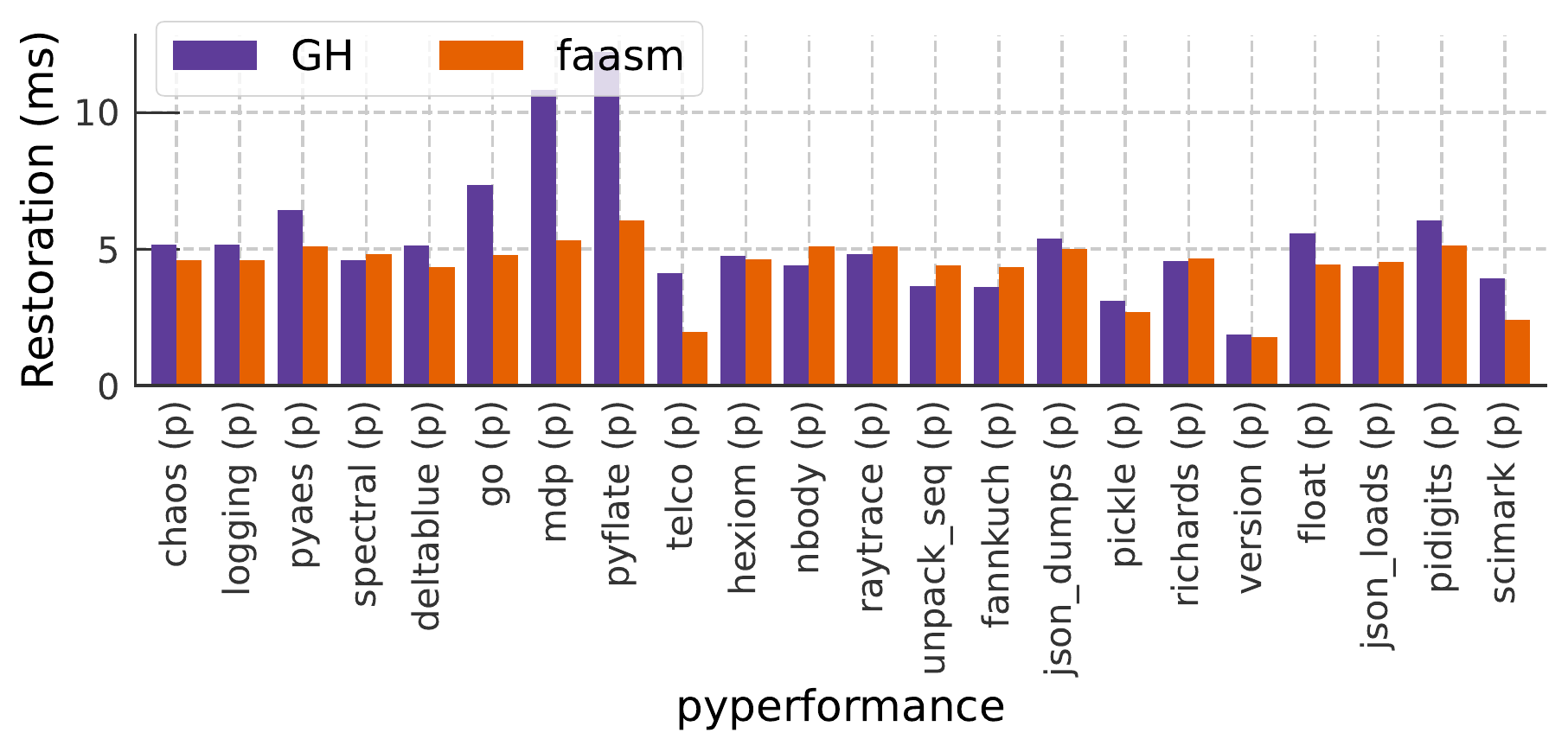}\\%
	\vspace*{-0.1cm}%
	\includegraphics[width=0.95\columnwidth]{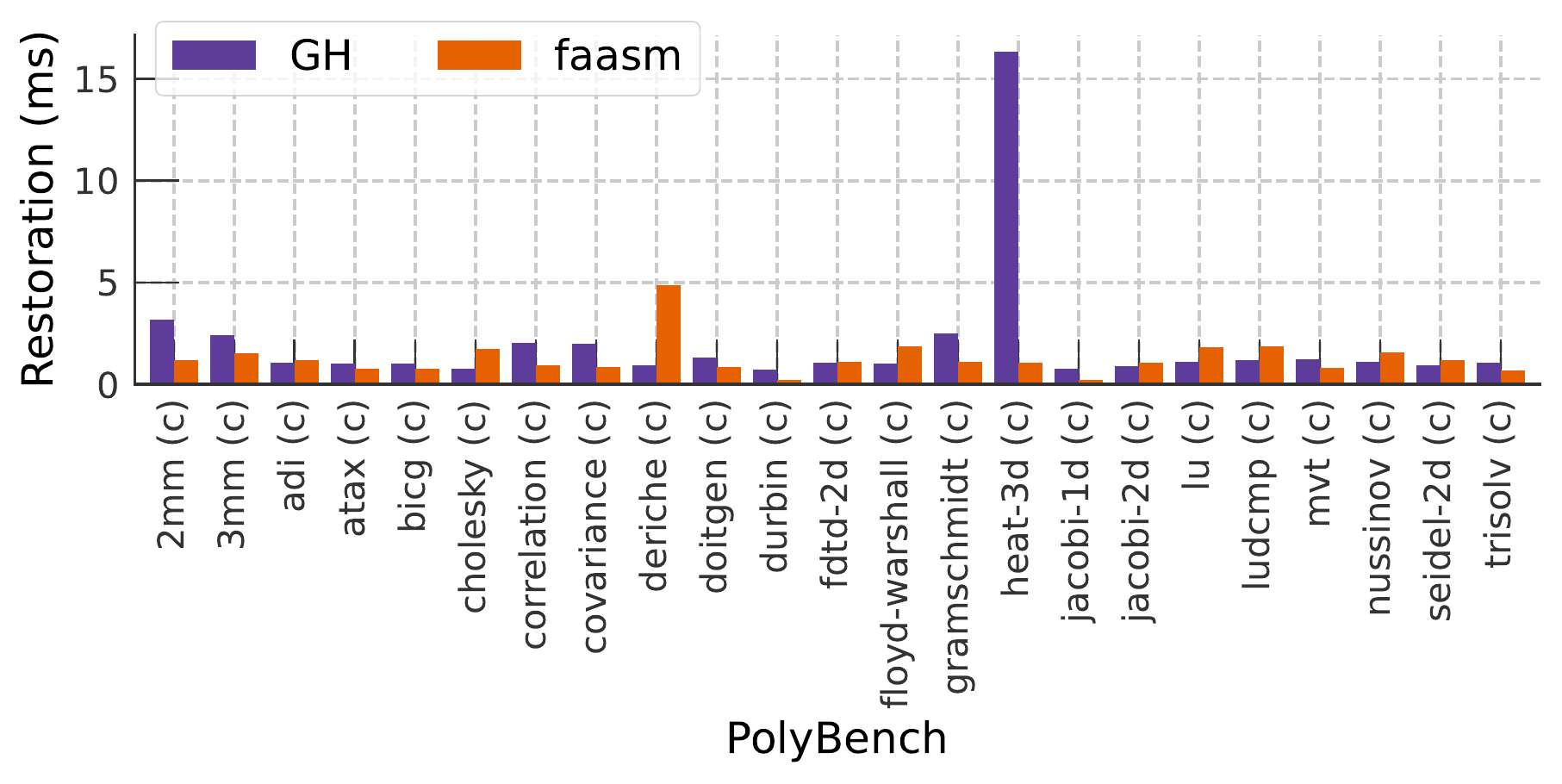}\\%
	\caption{Restoration duration (off the critical pathe) of \lsys, and \lfaasm. The symbols (p) and (c) denote Python and C.}%
	\label{fig:reset}%
\end{figure}

\autoref{fig:latency} shows latencies for \lfaasmshaded next to those
for \lsys.  On most pyperformance benchmarks, \lfaasm latency is
considerably higher than that of \lsys, whereas the restoration time
is comparable \autoref{fig:reset}.  This is because the Python
interpreter and runtime are less efficient when compiled to {\Wasm}
(which \faasm uses) compared to a natively compiled interpreter (which
\lsys uses).

On PolyBench functions, {\lfaasm} latencies are
generally lower than those of {\lsys}. However, {\lsys}'s poorer
relative performance are not because of {\sys}'s overheads. Rather,
{\Wasm}'s runtime is specifically optimized for program patterns
that occur in PolyBench, so {\Wasm} compiled PolyBench outperforms
natively compiled PolyBench even in the baseline (this observation has been noted
in prior work~\cite{faasm,haas2017bringing,jangda2019not}).

\update{The same trends continue to manifest in throughput
  measurements where \lfaasm has lower throughput than \lsys on most
  pyperformance functions, and higher throughput than than \lsys on
  most PolyBench functions. We omit the detailed throughput comparison
  here as it entangles many variables such as the
  differences in the platforms, which have nothing to do with request
  isolation, the platforms' internal components, runtimes (native vs
  \Wasm), as well as the isolation mechanisms. The reader can find
  these numbers in \autoref{appendix:grandtables}.%

Overall, the performance differences between \lfaasm and \lsys are
dominated by differences between native and {\Wasm} compilation rather
than request isolation costs.}

\subsubsection{Throughput scaling with cores}~
\label{sec:eval:throughputscalability}

\begin{figure*}[htp]
	\centering
	\includegraphics[width={\textwidth}]{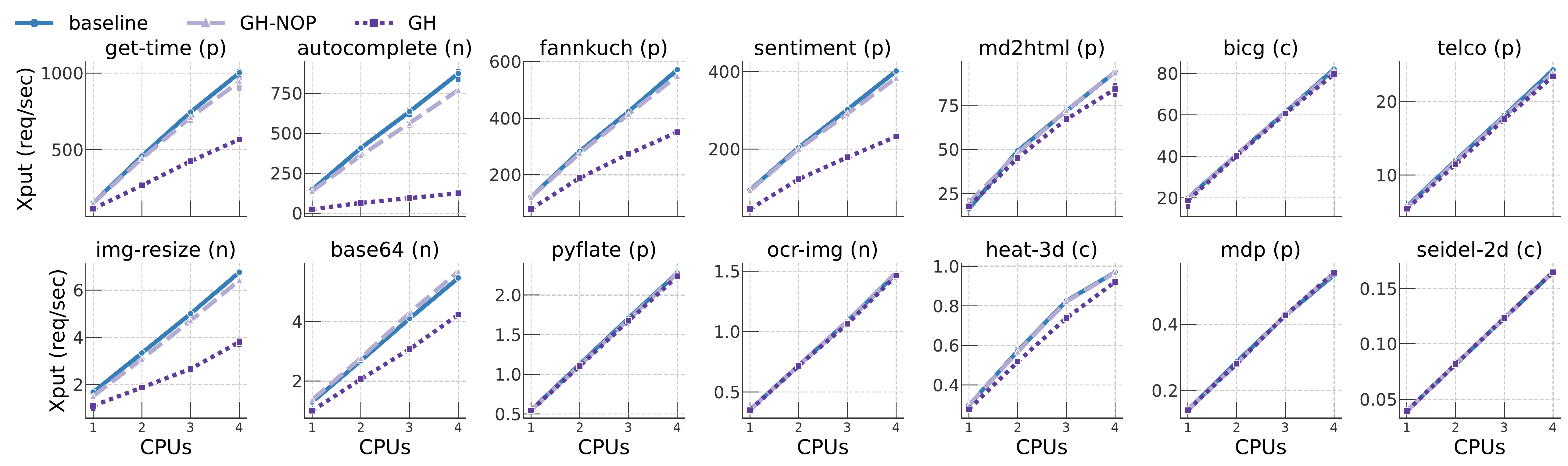}
	\sfigcap{Throughput scaling with number of cores on a subset of benchmarks}\label{fig:scalability-gh}
\end{figure*}%

We expect {\lsys}'s throughput to scale linearly with cores as each
core can run a completely independent container instance with its own
function and {\sys} copy. To confirm this, we repeat the throughput
experiment above, varying the number of cores available to the VM from
1 to 4 (and an equal number of function container
instances\finalpass{, each limited to 1 core}). \autoref{fig:scalability-gh} shows absolute throughputs as
a function of the number of available cores for a subset of 14
representative benchmarks of varying duration, number of mapped pages
and number of dirtied pages. Reported numbers are sustained
throughputs averaged over 6 runs of at least 1.5 minutes each
(excluding a warm-up).  Error bars are standard deviations over the 6
runs. As expected, the scaling is nearly linear in all
cases. \resolvepeter{We expect this nearly-linear trend to continue
  beyond 4 cores until a bottleneck in the kernel or memory buses
  arises.}

\subsection{Deconstructing restoration overheads}
\label{eval:restoration}
\begin{figure}[htb]%
	\centering%
	\includegraphics[clip, trim=0.3cm 0.4cm 0.5cm 0.3cm,width=\columnwidth]{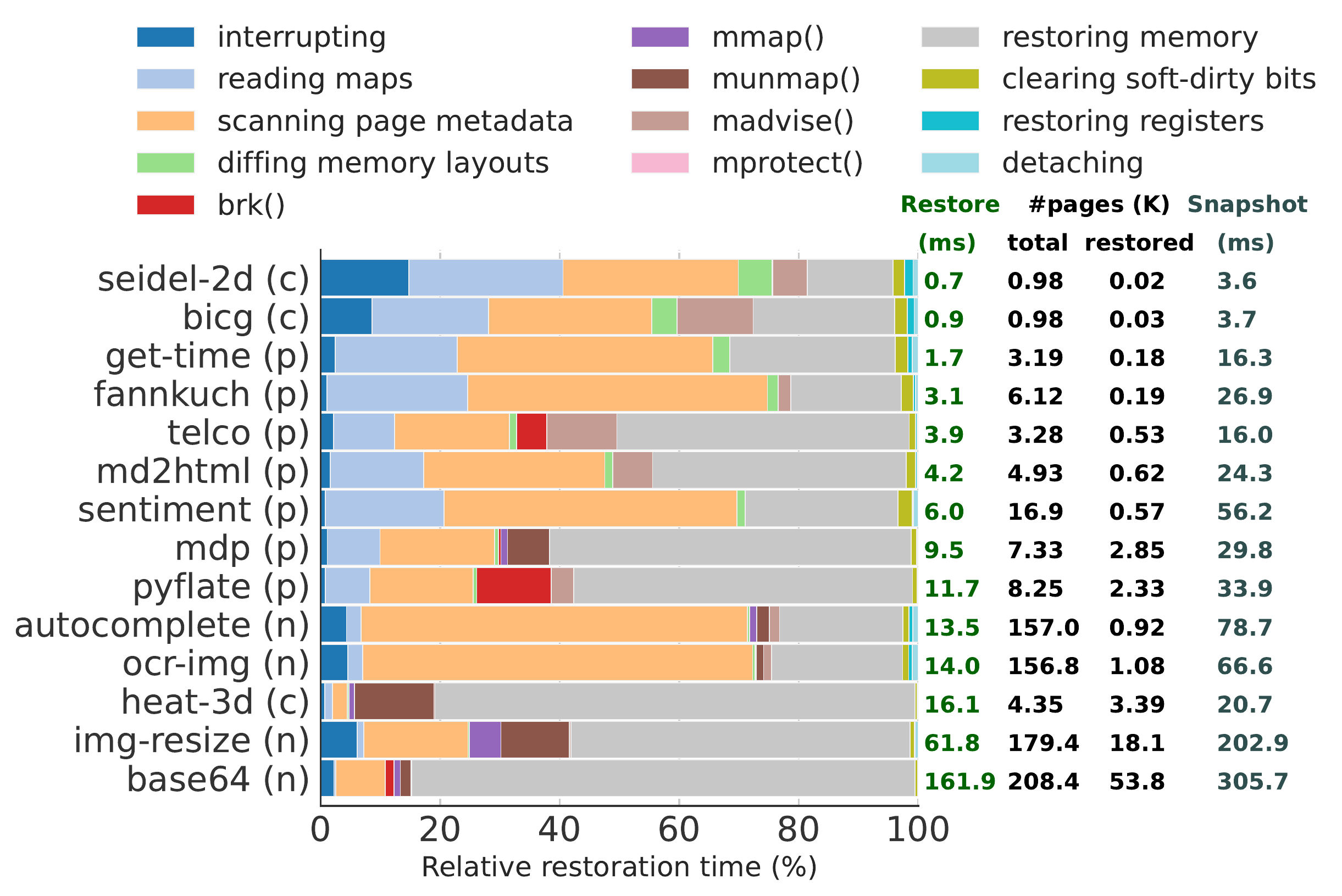}%
	\figcap{Restoration overhead (deconstructed) and the one-time snapshotting overhead for a subset of benchmarks}%
	\label{fig:restoration_overheads}%
\end{figure}

\newcommand{\coloredbox}[1]{%
\raisebox{0mm}{\begin{tikzpicture}[
  x=\textwidth/100,y=\textwidth/100,
  every node/.style={inner ysep=0.2,outer sep=0, inner xsep=1},
  ]
\node [minimum height=2mm, minimum width=4mm, fill=#1, draw=none] at (0,0) {};
\end{tikzpicture}}}
\newcommand{\coloredboxthin}[1]{%
\raisebox{0mm}{\begin{tikzpicture}[
  x=\textwidth/100,y=\textwidth/100,
  every node/.style={inner ysep=0.2,outer sep=0, inner xsep=1},
  ]
\node [minimum height=2mm, minimum width=0.8mm, fill=#1, draw=none] at (0,0) {};
\end{tikzpicture}}}

\definecolor{interrupting}{HTML}{1f77b4}
\definecolor{reading}{HTML}{aec7e8}
\definecolor{scanning}{HTML}{ffbb78}
\definecolor{diffing}{HTML}{98df8a}
\definecolor{brk}{HTML}{d62728}
\definecolor{mmap}{HTML}{9467bd}
\definecolor{munmap}{HTML}{8c564b}
\definecolor{madvise}{HTML}{c49c94}
\definecolor{mprotect}{HTML}{f7b6d2}
\definecolor{restoring}{HTML}{c7c7c7}
\definecolor{clearing}{HTML}{bcbd22}
\definecolor{registers}{HTML}{17becf}
\definecolor{detaching}{HTML}{9edae5}

\sys restoration involves several steps that we outlined
in~\autoref{subsec:restoration}.  In this section we break down the
cost of restoration for the same 14 representative benchmarks %
(selection criteria in~\autoref{sec:eval:throughputscalability}).  The overall restoration
cost breaks down into the following components:
\noindent\begin{itemize}[noitemsep,topsep=0pt,parsep=0pt,partopsep=0pt,label={},leftmargin=0.75cm]
\item[\coloredbox{interrupting}] interrupting the function process.
\item[\coloredbox{reading}] reading the process' memory mapped regions
\item[\coloredbox{scanning}] scanning all mapped pages to identify which are dirtied
\item[\coloredbox{diffing}] diffing the memory layout to identify how it has changed
\item[\coloredboxthin{brk}\coloredboxthin{mmap}\coloredboxthin{munmap}\coloredboxthin{madvise}\coloredboxthin{mprotect}] restoring the original memory layout by injecting syscalls (brk, mmap, munmap, madvise, and mprotect)
\item[\coloredbox{restoring}] restoring the contents of modified and removed pages
\item[\coloredbox{clearing}] restoring registers
\item[\coloredbox{registers}] resetting the soft-dirty bits of all modified pages
\item[\coloredbox{detaching}] detaching from the process
\end{itemize}

Each of these costs depends on different factors.  The costs of interrupting, restoring registers, and detaching are functions of the number of threads in the process.  The costs of reading, scanning, diffing \finalpass{the memory layout}, and resetting soft-dirty bits, are functions of the address space size\finalpass{~and layout}.  The syscall injection cost depends on the number of memory layout changes and is heavily dependent on the language runtime.  Lastly, the cost of restoring the contents of pages depends on the number of pages dirtied during an invocation.

\autoref{fig:restoration_overheads} shows these costs normalized to
the total restoration cost for our 14 representative functions.  For
each benchmark we also detail the absolute restoration time, the
number of pages, and the time for \sys to take its initial snapshot
(we revisit snapshotting overhead
in~\autoref{subsec:snapshotting}). \resolvepeter{In particular, we
  note that the \finalpass{memory} restoration cost (\coloredbox{restoring}) is strongly
  correlated with the total number of pages restored.  Similarly, the
  time spent scanning page metadata (\coloredbox{scanning}) is
  strongly correlated with the total number of pages (as discussed in
  ~\autoref{sec:design:tracking}, optimizations can make the costs
  correlate to the number of dirtied pages instead).}

\subsection{Snapshotting overhead}
\label{subsec:snapshotting}

The rightmost column of \autoref{fig:restoration_overheads} outlines
\sys's snapshotting latency overhead for the same 14 functions that we
used in \autoref{sec:eval:throughputscalability}.  Recall that
snapshotting is a one-time operation that occurs upon container
initialization.  It involves warming the function by making a dummy
request, pausing the process, and copying the process's state to
\sys's manager process memory. \resolvepeter{Snapshotting time and memory costs are primarily
  proportional to the total number of paged memory pages. The
  snapshotting latency overhead can be alleviated using the same
  techniques that reduce cold start latencies (Catalyzer
  \cite{catalyzer}, REAP \cite{REAP}, FaaSnap \cite{faasnap},
  Replayable \cite{wang2019replayable}, Prebaking \cite{prebaking},
  Pagurus \cite{pagurus}) by checkpointing the initialized \sys
  process along with the function's process. \sys memory overhead
  could be easily reduced to be proportional to the number of dirtied
  memory pages at the cost of a one-time on-critical-path
  copy-on-write per unique modified page in the function's
  life-cycle. Since snapshotting is an infrequent operation in \sys,
  we have not attempted these optimizations.}

\section{Related work}
\label{sec:related}
\label{subsec:related-cold-starts}

\paragraph{Fork-based request isolation}
A standard technique for request isolation in services, not FaaS
specifically, is to fork a clean state to serve every request. For
example, the Apache web server \cite{apache-prefork}\footnote{\finalpass{Using the default Apache Prefork MPM.}} uses this approach to isolate client
sessions from each other. The same idea can be used for request
isolation in FaaS. However, {\tt fork()} does not work with multithreaded
functions or runtimes without extensive modifications to prepare all
threads for a consistent snapshot~\cite{catalyzer}. Even for
single-threaded functions, a fork-based approach is less performant
than {\sys} (see \autoref{sec:evaluation}) due to the high cost of the
fork syscall and page-copying faults on the critical path for all
written pages. The cost of fork itself can be reduced using lighter
process-like abstractions such as lightweight contexts
(lwCs)~\cite{litton2016light}, but this does not reduce the cost of
page copying on the critical path.

\paragraph{Advances in reducing container cold-start latencies}
Reducing container cold-start latencies is an active area of research. Several
techniques have been proposed, including maintaining pre-warmed idle
containers for a
function~\cite{aws-provisioned-concurrency,ow-prewarm}, maintaining a
pool of containers that can be
repurposed~\cite{mohan2019agile,Stenbom2019refunction}, maintaining
partially initialized runtimes with loaded libraries as in
SOCK~\cite{oakes2018sock}, relaxing isolation between functions by
allowing functions from the same app developer to share containers
(SAND~\cite{akkus2018sand}, Azure~\cite{azure-functions}), and
starting from slim container images and adding non-essential functions
only when needed (CNTR~\cite{thalheim2018cntr}). These techniques do
not provide request isolation, the problem that {\sys} targets, but
they can be combined with {\sys} to solve the cold-start latency and
the request isolation problems simultaneously.

Other methods of reducing cold-start latencies rely on snapshotting
and restoration, which {\sys} also
uses. Replayable \cite{wang2019replayable} noted the phased nature of
runtime initialization, so snapshotting after (part of) the
initialization phase, and starting cold invocations from such a
snapshot lowers cold-start latencies. In principle, this approach can
also be used for request isolation by starting each invocation from
such a snapshot but previous snapshot/restore techniques
have overheads that are orders of magnitude higher than those of {\sys}.

Snapshotting techniques based on
CRIU~\cite{CRIU,Chen2015microservicesCRIU,vasavada2011comparing,cooperman2006transparent,rieker2006transparent}
serialize snapshots to persistent storage and are insufficient for
request isolation due to the high overhead of deserialization during
restoration, which is on the order of seconds.

Techniques that store snapshots in memory lower this overhead, but not
sufficiently. For example, VAS-CRIU~\cite{venkatesh2019fast} treats
the address spaces as a first-class OS primitive, allowing an address
space to be attached to any process. However, container restoration
time is still of the order of \textasciitilde0.5s.
SEUSS~\cite{cadden2020seuss} takes a unikernel approach, building a
customized VM for each function where everything runs in kernel
space. SEUSS allows incremental snapshots to jump-start
functions. However, SEUSS (and VAS-CRIU) rely on copy-on-write, thus
increasing the in-function latency, like the fork-based approach.

Catalyzer~\cite{catalyzer} trades function-start latency for
in-function latency using a lazy restoration that incurs page
faults. REAP \cite{REAP} reduces the cost of these page faults by
eagerly pre-fetching pages that were part of the active working set of
the function in the past. However, overall function latencies after a
restoration are still high: For a simple hello-world function that
executes in 1ms without restoration, Catalyzer and REAP latencies with
restoration are 232ms and 60ms, respectively. In contrast, \sys can
restore a C hello world function in \textasciitilde\ums{0.5} and an
equivalent Python function in \textasciitilde{}\ums{1.7} off the
critical path.

\resolvepeter{FaaSnap \cite{faasnap} performs a different optimization
  -- it enhances the pre-fetching of pages. For instance, it does
  concurrent prefetching while the VM is loading, and fetches pages in
  the approximate order of loading such that pages have a higher
  chance of being fetched by the time the function needs them. These
  optimizations further reduce the latency of cold-starts by 1.4x
  relative to a baseline without the optimizations. Nonetheless,
  overheads are high: The restoration of a simple hello world in
  FaaSnap takes as much time as it does in REAP.}

Cloudflare Workers~\cite{cloudflare}, Faastly~\cite{faastly,lucet},
and \faasm~\cite{faasm} solve the cold-start problem by relying on
software-fault isolation (SFI) using V8 isolates and
{\wasm}~\cite{wasm}. Here, several function spaces -- called
Faaslets in \faasm -- are packed into a single running process,
relying on SFI to isolate them from each other. Obtaining a fresh
Faaslet for a function invocation amounts to remapping an unused
Faaslet's heap to a previously checkpointed, pre-warmed state of the
function. {\wasm} limits the heap to a contiguous 4GB region, so
this remapping is fast and effectively solves the cold-start
problem. The \faasm paper notes that the same idea can be used for
efficient request isolation by applying the remapping between
requests. We compared the performance of this request isolation
approach to that of {\sys} in \autoref{sec:evaluation} and showed that
the trade-offs are very different. Unlike {\sys}, this technique is
limited to languages, runtimes and threading models that can be
compiled to {\wasm}.

\section{Conclusion}
\label{sec:conclusion}
\finalpass{
\sys relies on efficient in-memory process state snapshot and restore to provide
sequential request isolation in FaaS platforms.  \sys's design is
agnostic to the FaaS platform, OS kernel, programming languages, runtimes,
and libraries used to write functions. \sys overheads on end-to-end
latency and throughput are modest, and lower than what could be achieved by repurposing state-of-the art techniques for solving the container cold-start problem to provide sequential request isolation.
}

\begin{acks}
We thank Antoine Kaufmann, Akram El-Korashy, Vaastav Anand, Anjo Vahldiek-Oberwagner, Aastha Mehta, and the anonymous reviewers for their helpful feedback on earlier versions of this
work. We thank Simon Shillaker for his help with reproducing \faasm results. This work was supported in part by the European
Research Council (ERC Synergy imPACT 610150) and the
German Science Foundation (DFG CRC 1223).
\end{acks}
\bibliographystyle{acm}
\bibliography{references}
\appendix
\clearpage

\begin{table*}%
	\section{Detailed evaluation results}
	\label{appendix:grandtables}

	\definecolor{mygray}{HTML}{CCCCCC}
\definecolor{mygreen}{HTML}{B7E1CD}
\definecolor{myred}{HTML}{E6B8AF}
\definecolor{mypurple}{HTML}{e4b6f0}
\definecolor{myblue}{HTML}{C9DAF8}
\definecolor{verylightgray}{HTML}{DDDDDD}

\colorlet{baselinenegative}{white}
\colorlet{baselinemiddle}{white}
\colorlet{baselinepositive}{white}

\colorlet{ghnegative}{myred}
\colorlet{ghmiddle}{white}
\colorlet{ghpositive}{myblue}

\colorlet{ghnopnegative}{myred}
\colorlet{ghnopmiddle}{white}
\colorlet{ghnoppositive}{myblue}

\colorlet{faasmnegative}{mygreen}
\colorlet{faasmmiddle}{white}
\colorlet{faasmpositive}{myred}

\colorlet{forknegative}{mygreen}
\colorlet{forkmiddle}{white}
\colorlet{forkpositive}{myred}

\newcommand{\baselinelabel}{\textbf{base}}
\newcommand{\groundhoglabel}{\textbf{Gh}}
\newcommand{\faasmlabel}{\textls[-75]{\textbf{faasm}}}
\newcommand{\forklabel}{\textbf{fork}}
\newcommand{\ghnooplabel}{\textbf{Gh$_{\hspace{-0.5mm}\text{\relsize{-1}{\textls[-100]{nop}}}}$}}

\newcommand{\latencylabel}{\footnotesize \shortstack{E2E\\Latency\\[-1mm](ms)}}
\newcommand{\invokerlabel}{\footnotesize \shortstack{Invoker\\Latency\\[-1mm](ms)}}
\newcommand{\tputlabel}{\footnotesize \shortstack{T'put\\[-1mm](req/s)}}

\newcommand{\greenshading}{%
\raisebox{-0.2mm}{\begin{tikzpicture}[
  x=\textwidth/100,y=\textwidth/100,
  every node/.style={inner ysep=0.2,outer sep=0, inner xsep=1},
  ]
\node [fill=mygreen, draw=mygray] at (0,0) {Green};
\end{tikzpicture}}\xspace}

\newcommand{\redshading}{%
\raisebox{-0.2mm}{\begin{tikzpicture}[
  x=\textwidth/100,y=\textwidth/100,
  every node/.style={inner ysep=0.2mm,outer sep=0, inner xsep=1},
  ]
\node [fill=myred, draw=mygray] at (0,0) {Red};
\end{tikzpicture}}\xspace}

\newcommand{\purpleshading}{%
\raisebox{-0.8mm}{\begin{tikzpicture}[
  x=\textwidth/100,y=\textwidth/100,
  every node/.style={inner ysep=0.2,outer sep=0, inner xsep=1},
  ]
\node [fill=mypurple, draw=mygray] at (0,0) {Purple};
\end{tikzpicture}}\xspace}

\newcommand{\blueshading}{%
\raisebox{-0.2mm}{\begin{tikzpicture}[
  x=\textwidth/100,y=\textwidth/100,
  every node/.style={inner ysep=0.2,outer sep=0, inner xsep=1},
  ]
\node [fill=myblue, draw=mygray] at (0,0) {Blue};
\end{tikzpicture}}\xspace}
	\includegraphics[trim=0 88 10 96,clip, width=\textwidth]{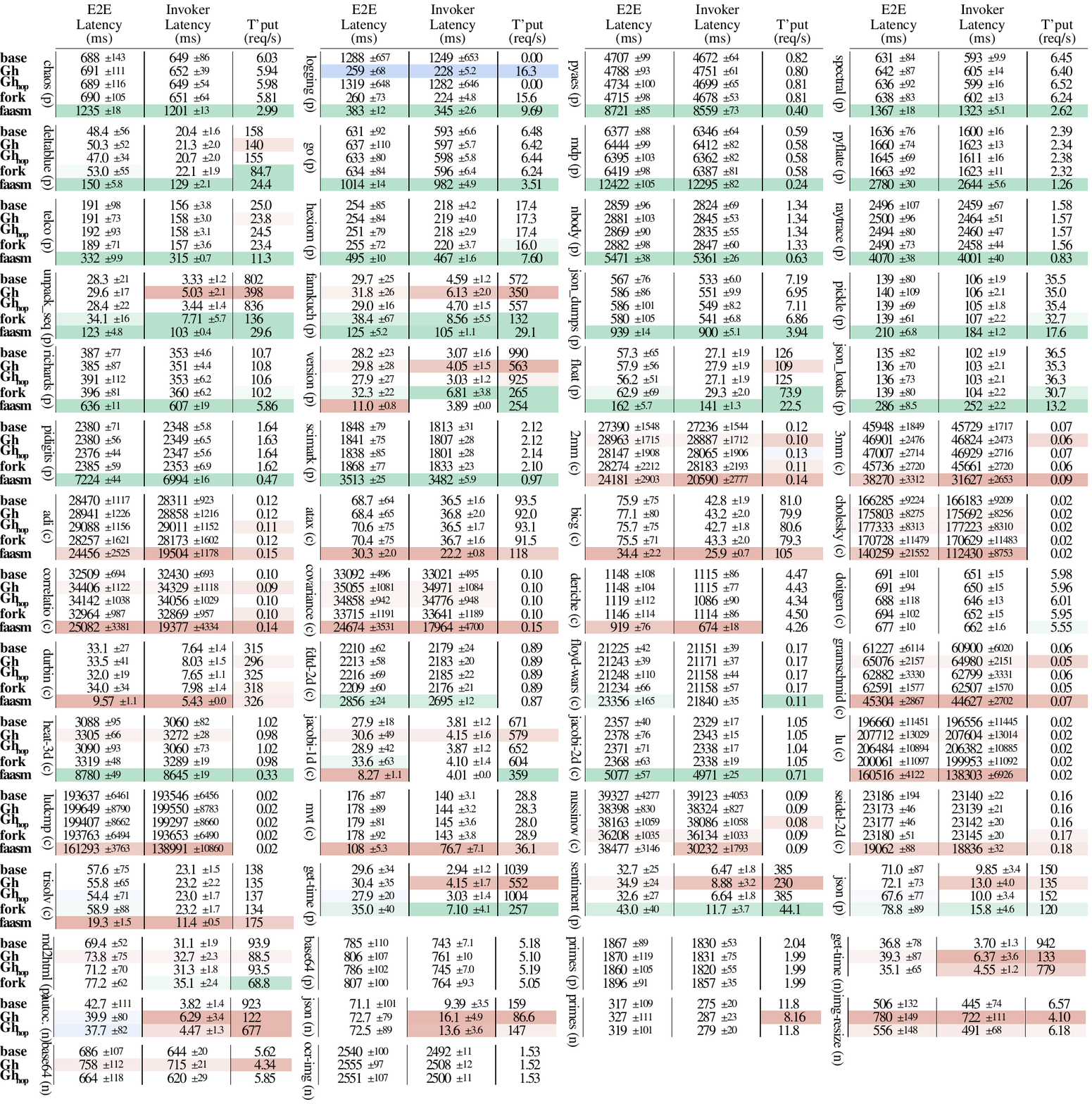}%
	\caption{Latency and throughput measurements comparing \sys to several other systems: \baselinelabel\xspace baseline OpenWhisk; \groundhoglabel\xspace \sys on OpenWhisk; \faasmlabel\xspace  faasm; \forklabel\xspace  fork-based implementation on OpenWhisk; and \ghnooplabel\xspace  \sys with no restoration.   We run 58 benchmarks across three languages indicated by \textbf{(p)} python, \textbf{(c)} c, and \textbf{(n)} \nodejs. We highlight cells of interest if there is more than a 5\% difference as follows: \protect\greenshading indicates results faorable to \sys \protect\redshading indicates results unfaorable to \sys \protect\blueshading indicates the unexpected result that \sys is outperforming the baseline. \textit{\textbf{Disclaimer}}: Faasm throughput measurements are provided for the curious reader, but no conclusions should be drawn based on them as they entangle many variables such as the difference in the platforms, their internal components and deployment, runtimes (native vs \Wasm), as well as the isolation mechanism}
	\label{tab:macrobenchmarks}

\end{table*}%

\thispagestyle{empty}
\begin{table*}[t]

	\centering
	\resizebox{\textwidth}{!}{
			\begin{tabular}{lccccc|ccccc|cc|cc}
		\toprule
		{\multirow{2}{0.5cm}{\thead{Benchmark}}}&
		\multicolumn{5}{c}{\thead{Base End-to-End Latency (ms) / rel. overhead (\%) ±CoV(\%)}}&
		\multicolumn{5}{c}{\thead{Throughput (r/s) / rel. overhead (\%)}}&
		\multicolumn{2}{c}{\thead{Inv. lat  (ms)/rel. (\%) ± CoV (\%)}}&
		\multicolumn{2}{c}{\thead{Restoration}}\\
		& baseline & GH-NOP & GH & fork & faasm
		& baseline & GH-NOP & GH & fork & faasm
		& baseline & GH & time (ms) & \#Kpages (\%))\\
	\midrule
\textbf{chaos (p)}          & 688.18±20.8           & +0.12 ±16.8                           & +0.36 ±16.1                          & +0.22 ±15.3                           & \cellcolor[HTML]{B7E1CD}+79.45 ±1.4  & 6.03                                         & -0.90                                            & -1.40                                            & -3.70                                  & {\cellcolor[HTML]{B7E1CD}-50.40} & 648.52±13.3             & +0.54 ±6.0                           & 4.9                                      & 0.47 (7.4)                      \\
\textbf{logging (p)}        & 1287.57±51.0          & +2.43 ±49.1                           & \cellcolor[HTML]{C9DAF8}-79.89 ±26.1 & -79.82 ±28.3                          & -70.26 ±3.1                          & 0                                            & 0                                                  & {\cellcolor[HTML]{C9DAF8}+inf} & {+inf}               & +inf                                               & 1249.39±52.2            & \cellcolor[HTML]{C9DAF8}-81.76 ±2.3  & 4.8                                      & 0.41 (6.7)                      \\
\textbf{pyaes (p)}          & 4707.26±2.1           & +0.57 ±2.1                            & +1.71 ±1.9                           & +0.16 ±2.1                            & \cellcolor[HTML]{B7E1CD}+85.26 ±1.0  & 0.82                                         & -0.40                                            & -1.30                                            & -1.00                                  & {\cellcolor[HTML]{B7E1CD}-50.80} & 4671.98±1.4             & +1.70 ±1.3                           & 6                                        & 0.84 (13.5)                     \\
\textbf{spectral (p)}       & 630.84±13.4           & +0.87 ±14.5                           & +1.84 ±13.6                          & +1.20 ±13.0                           & \cellcolor[HTML]{B7E1CD}+116.72 ±1.3 & 6.45                                         & {+1.1}                         & -0.80                                            & -3.20                                  & {\cellcolor[HTML]{B7E1CD}-59.30} & 592.76±1.7              & +2.09 ±2.3                           & 4.3                                      & 0.21 (3.4)                      \\
\textbf{deltablue (p)}      & 48.44±116.6           & -2.97 ±71.6                           & +3.88 ±102.6                         & \cellcolor[HTML]{B7E1CD}+9.51 ±103.8  & \cellcolor[HTML]{B7E1CD}+210.47 ±3.8 & 157.63                                       & -1.70                                            & \cellcolor[HTML]{E6B8AF}-11.00                   & \cellcolor[HTML]{B7E1CD}-46.30         & {\cellcolor[HTML]{B7E1CD}-84.50} & 20.43±7.9               & +4.33 ±9.5                           & 4.6                                      & 0.33 (5.3)                      \\
\textbf{go (p)}             & 631.16±14.5           & +0.32 ±12.7                           & +0.91 ±17.3                          & +0.41 ±13.2                           & \cellcolor[HTML]{B7E1CD}+60.67 ±1.4  & 6.48                                         & -0.60                                            & -0.90                                            & -3.60                                  & {\cellcolor[HTML]{B7E1CD}-45.80} & 592.99±1.1              & +0.61 ±1.0                           & 6.9                                      & 0.95 (15.2)                     \\
\textbf{mdp (p)}            & 6377.46±1.4           & +0.27 ±1.6                            & +1.05 ±1.5                           & +0.65 ±1.5                            & \cellcolor[HTML]{B7E1CD}+94.77 ±0.8  & 0.59                                         & -0.60                                            & -1.00                                            & -1.40                                  & {\cellcolor[HTML]{B7E1CD}-58.80} & 6345.53±1.0             & +1.05 ±1.3                           & 9.6                                      & 2.85 (38.9)                     \\
\textbf{pyflate (p)}        & 1635.93±4.7           & +0.53 ±4.2                            & +1.45 ±4.5                           & +1.65 ±5.5                            & \cellcolor[HTML]{B7E1CD}+69.92 ±1.1  & 2.39                                         & -0.60                                            & -2.10                                            & -3.20                                  & {\cellcolor[HTML]{B7E1CD}-47.30} & 1599.84±1.0             & +1.42 ±0.8                           & 11.7                                     & 2.33 (28.2)                     \\
\textbf{telco (p)}          & 190.81±51.5           & +0.80 ±48.2                           & +0.30 ±38.2                          & -0.83 ±37.3                           & \cellcolor[HTML]{B7E1CD}+74.13 ±3.0  & 25.01                                        & -2.00                                            & \cellcolor[HTML]{E6B8AF}-4.90                    & \cellcolor[HTML]{B7E1CD}-6.50          & {\cellcolor[HTML]{B7E1CD}-54.70} & 155.64±2.4              & +1.54 ±1.9                           & 3.9                                      & 0.53 (16.2)                     \\
\textbf{hexiom (p)}         & 253.92±33.5           & -1.18 ±31.6                           & +0.13 ±33.2                          & +0.58 ±28.0                           & \cellcolor[HTML]{B7E1CD}+94.80 ±2.1  & 17.45                                        & -0.10                                            & -1.00                                            & \cellcolor[HTML]{B7E1CD}-8.30          & {\cellcolor[HTML]{B7E1CD}-56.50} & 218.21±1.9              & +0.45 ±1.8                           & 4.3                                      & 0.28 (4.5)                      \\
\textbf{nbody (p)}          & 2858.53±3.4           & +0.38 ±3.1                            & +0.78 ±3.6                           & +0.83 ±3.4                            & \cellcolor[HTML]{B7E1CD}+91.39 ±0.7  & 1.34                                         & {+0.0}                         & -0.10                                            & -0.70                                  & {\cellcolor[HTML]{B7E1CD}-53.20} & 2823.65±2.4             & +0.76 ±1.9                           & 4.1                                      & 0.21 (3.4)                      \\
\textbf{raytrace (p)}       & 2495.66±4.3           & -0.06 ±3.2                            & +0.17 ±3.8                           & -0.22 ±2.9                            & \cellcolor[HTML]{B7E1CD}+63.10 ±0.9  & 1.58                                         & -0.30                                            & -0.60                                            & -1.10                                  & {\cellcolor[HTML]{B7E1CD}-47.60} & 2459.2±2.7              & +0.19 ±2.1                           & 4.4                                      & 0.35 (5.6)                      \\
\textbf{unpack\_seq (p)}    & 28.34±74.3            & +0.27 ±75.8                           & +4.45 ±56.3                          & \cellcolor[HTML]{B7E1CD}+20.19 ±47.5  & \cellcolor[HTML]{B7E1CD}+333.81 ±3.9 & 801.86                                       & {+4.3}                         & \cellcolor[HTML]{E6B8AF}-50.40                   & \cellcolor[HTML]{B7E1CD}-83.00         & {\cellcolor[HTML]{B7E1CD}-96.30} & 3.32±36.8               & \cellcolor[HTML]{E6B8AF}+51.21 ±40.9 & 3.2                                      & 0.2 (3.3)                       \\
\textbf{fannkuch (p)}       & 29.7±83.6             & -2.46 ±56.3                           & \cellcolor[HTML]{E6B8AF}+7.04 ±83.3  & \cellcolor[HTML]{B7E1CD}+29.26 ±173.9 & \cellcolor[HTML]{B7E1CD}+319.97 ±4.2 & 572.32                                       & -2.80                                            & \cellcolor[HTML]{E6B8AF}-38.80                   & \cellcolor[HTML]{B7E1CD}-77.00         & {\cellcolor[HTML]{B7E1CD}-94.90} & 4.59±27.1               & \cellcolor[HTML]{E6B8AF}+33.77 ±32.7 & 3.1                                      & 0.19 (3.1)                      \\
\textbf{json\_dumps (p)}    & 567.41±13.3           & +3.26 ±17.3                           & +3.19 ±14.6                          & +2.13 ±18.2                           & \cellcolor[HTML]{B7E1CD}+65.41 ±1.5  & 7.19                                         & -1.10                                            & -3.40                                            & \cellcolor[HTML]{B7E1CD}-4.70          & {\cellcolor[HTML]{B7E1CD}-45.30} & 533.09±1.1              & +3.45 ±1.8                           & 4.9                                      & 0.51 (8.0)                      \\
\textbf{pickle (p)}         & 139.26±57.7           & +0.01 ±49.5                           & +0.64 ±77.5                          & -0.07 ±43.6                           & \cellcolor[HTML]{B7E1CD}+50.49 ±3.2  & 35.49                                        & -0.20                                            & -1.40                                            & \cellcolor[HTML]{B7E1CD}-7.80          & {\cellcolor[HTML]{B7E1CD}-50.60} & 105.64±1.8              & +0.01 ±2.0                           & 2.9                                      & 0.23 (6.7)                      \\
\textbf{richards (p)}       & 387.47±19.9           & +1.01 ±28.7                           & -0.66 ±22.7                          & +2.22 ±20.4                           & \cellcolor[HTML]{B7E1CD}+64.22 ±1.7  & 10.68                                        & -0.50                                            & {+1.6}                         & \cellcolor[HTML]{B7E1CD}-4.10          & {\cellcolor[HTML]{B7E1CD}-45.20} & 353.13±1.3              & -0.56 ±1.3                           & 4.2                                      & 0.23 (3.7)                      \\
\textbf{version (p)}        & 28.24±82.1            & -1.35 ±98.6                           & \cellcolor[HTML]{E6B8AF}+5.33 ±92.7  & \cellcolor[HTML]{B7E1CD}+14.34 ±66.8  & \cellcolor[HTML]{E6B8AF}-61.13 ±7.1  & 990.38                                       & \cellcolor[HTML]{E6B8AF}-6.60                    & \cellcolor[HTML]{E6B8AF}-43.20                   & \cellcolor[HTML]{B7E1CD}-73.30         & {\cellcolor[HTML]{B7E1CD}-74.30} & 3.07±50.5               & \cellcolor[HTML]{E6B8AF}+31.82 ±36.0 & 1.7                                      & 0.17 (5.4)                      \\
\textbf{float (p)}          & 57.28±112.7           & -1.80 ±90.7                           & +1.07 ±96.4                          & \cellcolor[HTML]{B7E1CD}+9.73 ±110.0  & \cellcolor[HTML]{B7E1CD}+182.82 ±3.5 & 125.98                                       & -0.80                                            & \cellcolor[HTML]{E6B8AF}-13.40                   & \cellcolor[HTML]{B7E1CD}-41.30         & {\cellcolor[HTML]{B7E1CD}-82.10} & 27.06±7.1               & +2.91 ±6.7                           & 5                                        & 0.65 (10.4)                     \\
\textbf{json\_loads (p)}    & 135.04±60.7           & +0.87 ±53.4                           & +1.04 ±51.2                          & +3.03 ±57.3                           & \cellcolor[HTML]{B7E1CD}+111.67 ±3.0 & 36.46                                        & -0.50                                            & -3.20                                            & \cellcolor[HTML]{B7E1CD}-15.70         & {\cellcolor[HTML]{B7E1CD}-63.70} & 101.98±1.9              & +1.33 ±2.1                           & 4                                        & 0.22 (3.6)                      \\
\textbf{pidigits (p)}       & 2380.0±3.0            & -0.19 ±1.8                            & -0.01 ±2.4                           & +0.20 ±2.5                            & \cellcolor[HTML]{B7E1CD}+203.53 ±0.6 & 1.64                                         & -0.20                                            & -0.50                                            & -1.00                                  & {\cellcolor[HTML]{B7E1CD}-71.50} & 2347.55±0.2             & +0.07 ±0.3                           & 5.4                                      & 0.81 (13.2)                     \\
\textbf{scimark (p)}        & 1848.12±4.3           & -0.54 ±4.6                            & -0.37 ±4.1                           & +1.09 ±4.1                            & \cellcolor[HTML]{B7E1CD}+90.08 ±0.7  & 2.12                                         & {+0.7}                         & {+0.0}                         & -0.90                                  & {\cellcolor[HTML]{B7E1CD}-54.40} & 1812.64±1.7             & -0.34 ±1.6                           & 3.8                                      & 0.52 (16.0)                     \\
\cmidrule{1-15}
\textbf{2mm (c)}            & 27390.28±5.7          & +2.76 ±6.8                            & \cellcolor[HTML]{E6B8AF}+5.74 ±5.9   & +3.22 ±7.8                            & \cellcolor[HTML]{E6B8AF}-11.72 ±12.0 & 0.12                                         & {+0.6}                         & \cellcolor[HTML]{E6B8AF}-16.10                   & \cellcolor[HTML]{E6B8AF}-12.50         & \cellcolor[HTML]{E6B8AF}+10.7                      & 27236.21±5.7            & \cellcolor[HTML]{E6B8AF}+6.06 ±5.9   & 3.1                                      & 0.02 (2.0)                      \\
\textbf{3mm (c)}            & 45947.69±4.0          & +2.30 ±5.8                            & +2.07 ±5.3                           & -0.46 ±5.9                            & \cellcolor[HTML]{E6B8AF}-16.71 ±8.7  & 0.07                                         & -1.80                                            & \cellcolor[HTML]{E6B8AF}-8.80                    & -8.80                                  & \cellcolor[HTML]{E6B8AF}+23.4                      & 45729.02±3.8            & +2.40 ±5.3                           & 2.3                                      & 0.02 (2.0)                      \\
\textbf{adi (c)}            & 28470.33±3.9          & +2.17 ±4.0                            & +1.65 ±4.2                           & -0.75 ±5.7                            & \cellcolor[HTML]{E6B8AF}-14.10 ±10.3 & 0.12                                         & \cellcolor[HTML]{E6B8AF}-5.80                    & -2.60                                            & -2.80                                  & \cellcolor[HTML]{E6B8AF}+24.6                      & 28311.08±3.3            & +1.93 ±4.2                           & 0.8                                      & 0.02 (2.0)                      \\
\textbf{atax (c)}           & 68.72±92.5            & +2.78 ±105.9                          & -0.42 ±94.3                          & +2.52 ±106.4                          & \cellcolor[HTML]{E6B8AF}-55.96 ±6.5  & 93.55                                        & -0.50                                            & -1.70                                            & -2.20                                  & \cellcolor[HTML]{E6B8AF}+25.8                      & 36.45±4.4               & +1.01 ±5.4                           & 1                                        & 0.03 (3.1)                      \\
\textbf{bicg (c)}           & 75.89±98.7            & -0.20 ±98.5                           & +1.59 ±103.1                         & -0.45 ±93.9                           & \cellcolor[HTML]{E6B8AF}-54.72 ±6.3  & 81.05                                        & -0.50                                            & -1.40                                            & -2.10                                  & \cellcolor[HTML]{E6B8AF}+29.1                      & 42.78±4.4               & +0.89 ±4.7                           & 0.9                                      & 0.03 (3.1)                      \\
\textbf{cholesky (c)}       & 166284.84±5.5         & \cellcolor[HTML]{E6B8AF}+6.64 ±4.7    & \cellcolor[HTML]{E6B8AF}+5.72 ±4.7   & +2.67 ±6.7                            & \cellcolor[HTML]{E6B8AF}-15.65 ±15.4 & 0.02                                         & \cellcolor[HTML]{E6B8AF}-8.60                    & -1.00                                            & -2.60                                  & +0.4                                               & 166182.8±5.5            & \cellcolor[HTML]{E6B8AF}+5.72 ±4.7   & 0.6                                      & 0.02 (2.0)                      \\
\textbf{correlation (c)}    & 32508.82±2.1          & \cellcolor[HTML]{E6B8AF}+5.02 ±3.0    & \cellcolor[HTML]{E6B8AF}+5.84 ±3.3   & +1.40 ±3.0                            & \cellcolor[HTML]{E6B8AF}-22.85 ±13.5 & 0.1                                          & {+2.4}                         & \cellcolor[HTML]{E6B8AF}-6.20                    & \cellcolor[HTML]{E6B8AF}-5.90          & \cellcolor[HTML]{E6B8AF}+43.1                      & 32429.64±2.1            & \cellcolor[HTML]{E6B8AF}+5.86 ±3.3   & 2                                        & 0.02 (2.0)                      \\
\textbf{covariance (c)}     & 33092.13±1.5          & \cellcolor[HTML]{E6B8AF}+5.34 ±2.7    & \cellcolor[HTML]{E6B8AF}+5.93 ±3.1   & +1.88 ±3.5                            & \cellcolor[HTML]{E6B8AF}-25.44 ±14.3 & 0.1                                          & -3.80                                            & {+0.8}                         & {+0.4}               & \cellcolor[HTML]{E6B8AF}+48.3                      & 33020.56±1.5            & \cellcolor[HTML]{E6B8AF}+5.91 ±3.1   & 2                                        & 0.02 (2.0)                      \\
\textbf{deriche (c)}        & 1148.32±9.4           & -2.56 ±10.1                           & -0.01 ±9.0                           & -0.16 ±9.9                            & \cellcolor[HTML]{E6B8AF}-19.94 ±8.3  & 4.47                                         & -3.00                                            & -1.00                                            & {+0.6}               & {-4.70}                          & 1114.99±7.7             & +0.00 ±6.9                           & 0.8                                      & 0.02 (2.0)                      \\
\textbf{doitgen (c)}        & 691.08±14.7           & -0.44 ±17.1                           & -0.01 ±13.7                          & +0.45 ±14.8                           & -2.00 ±1.5                           & 5.98                                         & {+0.6}                         & -0.30                                            & -0.40                                  & {\cellcolor[HTML]{B7E1CD}-7.00}  & 650.53±2.2              & -0.08 ±2.3                           & 1.3                                      & 0.02 (2.0)                      \\
\textbf{durbin (c)}         & 33.1±82.8             & -3.25 ±59.1                           & +1.31 ±122.9                         & +2.64 ±100.6                          & \cellcolor[HTML]{E6B8AF}-71.09 ±11.8 & 314.68                                       & {+3.2}                         & \cellcolor[HTML]{E6B8AF}-6.00                    & {+1.0}               & \cellcolor[HTML]{E6B8AF}+3.8                       & 7.64±17.6               & \cellcolor[HTML]{E6B8AF}+5.05 ±18.4  & 0.6                                      & 0.02 (2.0)                      \\
\textbf{fdtd-2d (c)}        & 2209.61±2.8           & +0.29 ±3.1                            & +0.15 ±2.6                           & -0.05 ±2.7                            & \cellcolor[HTML]{B7E1CD}+29.23 ±0.9  & 0.89                                         & -0.50                                            & -0.30                                            & {+0.1}               & {-2.30}                          & 2179.15±1.1             & +0.16 ±0.9                           & 1                                        & 0.02 (2.0)                      \\
\textbf{floyd-warshall (c)} & 21224.8±0.2           & +0.11 ±0.5                            & +0.09 ±0.2                           & +0.05 ±0.3                            & \cellcolor[HTML]{B7E1CD}+10.04 ±0.7  & 0.17                                         & -0.30                                            & -1.80                                            & -1.20                                  & {\cellcolor[HTML]{B7E1CD}-35.20} & 21151.44±0.2            & +0.09 ±0.2                           & 0.8                                      & 0.02 (2.0)                      \\
\textbf{gramschmidt (c)}    & 61226.6±10.0          & +2.70 ±5.3                            & \cellcolor[HTML]{E6B8AF}+6.29 ±3.3   & +2.23 ±2.5                            & \cellcolor[HTML]{E6B8AF}-26.01 ±6.3  & 0.06                                         & -1.10                                            & \cellcolor[HTML]{E6B8AF}-11.10                   & \cellcolor[HTML]{E6B8AF}-6.20          & \cellcolor[HTML]{E6B8AF}+18.5                      & 60899.77±9.9            & \cellcolor[HTML]{E6B8AF}+6.70 ±3.3   & 2.5                                      & 0.02 (2.0)                      \\
\textbf{heat-3d (c)}        & 3088.12±3.1           & +0.07 ±3.0                            & \cellcolor[HTML]{E6B8AF}+7.02 ±2.0   & +7.47 ±1.5                            & \cellcolor[HTML]{B7E1CD}+184.32 ±0.6 & 1.02                                         & -0.40                                            & -4.00                                            & \cellcolor[HTML]{B7E1CD}-4.10          & {\cellcolor[HTML]{B7E1CD}-67.60} & 3059.55±2.7             & \cellcolor[HTML]{E6B8AF}+6.94 ±0.9   & 16.1                                     & 3.39 (77.9)                     \\
\textbf{jacobi-1d (c)}      & 27.92±63.9            & +3.34 ±144.5                          & \cellcolor[HTML]{E6B8AF}+9.65 ±159.4 & \cellcolor[HTML]{B7E1CD}+20.44 ±188.8 & \cellcolor[HTML]{E6B8AF}-70.39 ±12.8 & 671.34                                       & -2.80                                            & \cellcolor[HTML]{E6B8AF}-13.80                   & \cellcolor[HTML]{E6B8AF}-10.00         & {\cellcolor[HTML]{B7E1CD}-46.60} & 3.81±32.7               & \cellcolor[HTML]{E6B8AF}+9.01 ±39.4  & 0.6                                      & 0.02 (2.0)                      \\
\textbf{jacobi-2d (c)}      & 2356.66±1.7           & +0.62 ±3.0                            & +0.90 ±3.2                           & +0.49 ±2.7                            & \cellcolor[HTML]{B7E1CD}+115.44 ±1.1 & 1.05                                         & -0.20                                            & 0.00                                             & {+0.3}               & {\cellcolor[HTML]{B7E1CD}-32.10} & 2329.32±0.7             & +0.60 ±0.6                           & 0.7                                      & 0.02 (2.0)                      \\
\textbf{lu (c)}             & 196660.22±5.8         & \cellcolor[HTML]{E6B8AF}+5.00 ±5.3    & \cellcolor[HTML]{E6B8AF}+5.62 ±6.3   & +1.73 ±5.5                            & \cellcolor[HTML]{E6B8AF}-18.38 ±2.6  & 0.02                                         & {\cellcolor[HTML]{CFE2F3}+5.8} & {+2.2}                         & {+3.3}               & \cellcolor[HTML]{E6B8AF}+11.6                      & 196555.78±5.8           & \cellcolor[HTML]{E6B8AF}+5.62 ±6.3   & 0.7                                      & 0.02 (2.0)                      \\
\textbf{ludcmp (c)}         & 193637.44±3.3         & +2.98 ±4.3                            & +3.10 ±4.4                           & +0.06 ±3.4                            & \cellcolor[HTML]{E6B8AF}-16.70 ±2.3  & 0.02                                         & {+0.1}                         & {+7.3}                         & {+7.0}               & +6.9                                               & 193545.91±3.3           & +3.10 ±4.4                           & 1                                        & 0.02 (2.0)                      \\
\textbf{mvt (c)}            & 176.37±49.5           & +1.40 ±45.2                           & +0.89 ±49.9                          & +0.75 ±51.9                           & \cellcolor[HTML]{E6B8AF}-38.62 ±4.9  & 28.78                                        & -2.60                                            & -1.70                                            & {+0.4}               & \cellcolor[HTML]{E6B8AF}+25.3                      & 140.33±2.2              & +2.86 ±2.2                           & 1.2                                      & 0.03 (3.1)                      \\
\textbf{nussinov (c)}       & 39326.91±10.9         & -2.96 ±2.8                            & -2.36 ±2.2                           & \cellcolor[HTML]{E6B8AF}-7.93 ±2.9    & -2.16 ±8.2                           & 0.09                                         & \cellcolor[HTML]{E6B8AF}-6.40                    & -3.00                                            & -0.90                                  & \cellcolor[HTML]{E6B8AF}+2.0                       & 39122.65±10.4           & -2.04 ±2.2                           & 1                                        & 0.02 (2.0)                      \\
\textbf{seidel-2d (c)}      & 23186.15±0.8          & -0.04 ±0.2                            & -0.06 ±0.2                           & -0.02 ±0.2                            & \cellcolor[HTML]{E6B8AF}-17.79 ±0.5  & 0.16                                         & -0.50                                            & {+0.0}                         & {+0.6}               & \cellcolor[HTML]{E6B8AF}+9.7                       & 23140.14±0.1            & -0.01 ±0.1                           & 0.8                                      & 0.02 (2.0)                      \\
\textbf{trisolv (c)}        & 57.62±130.2           & \cellcolor[HTML]{CFE2F3}-5.65 ±130.4  & -3.14 ±115.6                         & +2.28 ±149.3                          & \cellcolor[HTML]{E6B8AF}-66.54 ±7.7  & 138.18                                       & -0.80                                            & -2.40                                            & -2.80                                  & \cellcolor[HTML]{E6B8AF}+26.6                      & 23.07±6.6               & +0.42 ±9.3                           & 1                                        & 0.02 (2.0)                      \\
\cmidrule{1-15}
\textbf{get-time (p)}       & 29.6±113.6            & \cellcolor[HTML]{CFE2F3}-5.68 ±72.6   & +2.71 ±114.8                         & \cellcolor[HTML]{B7E1CD}+18.13 ±114.1 &                                          & 1038.74                                      & -3.30                                            & \cellcolor[HTML]{E6B8AF}-46.90                   & \cellcolor[HTML]{B7E1CD}-75.30         &                                                      & 2.94±40.5               & \cellcolor[HTML]{E6B8AF}+41.08 ±41.0 & 1.7                                      & 0.18 (5.6)                      \\
\textbf{sentiment (p)}      & 32.67±76.3            & -0.10 ±82.0                           & \cellcolor[HTML]{E6B8AF}+6.66 ±68.5  & \cellcolor[HTML]{B7E1CD}+31.69 ±92.4  &                                          & 385.07                                       & 0.00                                             & \cellcolor[HTML]{E6B8AF}-40.20                   & \cellcolor[HTML]{B7E1CD}-88.50         &                                                      & 6.47±27.2               & \cellcolor[HTML]{E6B8AF}+37.26 ±35.8 & 6                                        & 0.57 (3.4)                      \\
\textbf{json (p)}           & 70.97±122.9           & \cellcolor[HTML]{CFE2F3}-4.81 ±113.7  & +1.56 ±101.8                         & \cellcolor[HTML]{B7E1CD}+11.04 ±113.0 &                                          & 150                                          & {+1.5}                         & \cellcolor[HTML]{E6B8AF}-9.80                    & \cellcolor[HTML]{B7E1CD}-19.80         &                                                      & 9.85±34.2               & \cellcolor[HTML]{E6B8AF}+31.68 ±30.6 & 3.7                                      & 0.87 (26.1)                     \\
\textbf{md2html (p)}        & 69.36±75.3            & +2.66 ±98.6                           & \cellcolor[HTML]{E6B8AF}+6.36 ±101.6 & \cellcolor[HTML]{B7E1CD}+11.38 ±80.0  &                                          & 93.94                                        & -0.50                                            & \cellcolor[HTML]{E6B8AF}-5.80                    & \cellcolor[HTML]{B7E1CD}-26.80         &                                                      & 31.04±6.3               & \cellcolor[HTML]{E6B8AF}+5.46 ±7.1   & 4.2                                      & 0.62 (12.6)                     \\
\textbf{base64 (p)}         & 785.33±14.0           & +0.03 ±13.0                           & +2.63 ±13.3                          & +2.81 ±12.4                           &                                          & 5.18                                         & {+0.2}                         & -1.50                                            & -2.50                                  &                                                      & 743.23±1.0              & +2.45 ±1.4                           & 7.7                                      & 1.66 (32.4)                     \\
\textbf{primes (p)}         & 1866.58±4.8           & -0.34 ±5.6                            & +0.17 ±6.3                           & +1.56 ±4.8                            &                                          & 2.04                                         & -2.40                                            & -2.30                                            & -2.40                                  &                                                      & 1829.74±2.9             & +0.05 ±4.1                           & 3.2                                      & 0.53 (16.5)                     \\
\cmidrule{1-15}
\textbf{get-time (n)}       & 36.84±211.2           & \cellcolor[HTML]{CFE2F3}-4.69 ±184.6  & \cellcolor[HTML]{E6B8AF}+6.68 ±220.4 &                                           &                                          & 942.07                                       & \cellcolor[HTML]{E6B8AF}-17.30                   & \cellcolor[HTML]{E6B8AF}-85.80                   & {}                     &                                                      & 3.7±34.8                & \cellcolor[HTML]{E6B8AF}+72.22 ±56.2 & 12.6                                     & 0.64 (0.4)                      \\
\textbf{autocomplete (n)}   & 42.74±260.6           & \cellcolor[HTML]{CFE2F3}-11.90 ±217.9 & \cellcolor[HTML]{C9DAF8}-6.71 ±201.2 &                                           &                                          & 922.59                                       & \cellcolor[HTML]{E6B8AF}-26.60                   & \cellcolor[HTML]{E6B8AF}-86.80                   & {}                     &                                                      & 3.82±36.8               & \cellcolor[HTML]{E6B8AF}+64.58 ±54.2 & 13.5                                     & 0.92 (0.6)                      \\
\textbf{json (n)}           & 71.1±142.4            & +1.99 ±122.7                          & +2.22 ±108.5                         &                                           &                                          & 159.09                                       & \cellcolor[HTML]{E6B8AF}-7.50                    & \cellcolor[HTML]{E6B8AF}-45.60                   & {}                     &                                                      & 9.4±37.8                & \cellcolor[HTML]{E6B8AF}+71.32 ±30.7 & 13                                       & 0.85 (0.5)                      \\
\textbf{primes (n)}         & 316.85±34.3           & +0.65 ±31.7                           & +3.27 ±33.9                          &                                           &                                          & 11.79                                        & {+0.1}                         & \cellcolor[HTML]{E6B8AF}-30.70                   & {}                     &                                                      & 274.63±7.3              & +4.56 ±8.0                           & 84.7                                     & 34.2 (17.0)                     \\
\textbf{img-resize (n)}     & 505.76±26.0           & \cellcolor[HTML]{E6B8AF}+9.87 ±26.6   & \cellcolor[HTML]{E6B8AF}+54.20 ±19.2 &                                           &                                          & 6.57                                         & \cellcolor[HTML]{E6B8AF}-6.00                    & \cellcolor[HTML]{E6B8AF}-37.60                   & {}                     &                                                      & 445.27±16.7             & \cellcolor[HTML]{E6B8AF}+62.09 ±15.3 & 61.8                                     & 18.05 (10.1)                    \\
\textbf{base64 (n)}         & 686.33±15.6           & -3.19 ±17.8                           & \cellcolor[HTML]{E6B8AF}+10.48 ±14.7 &                                           &                                          & 5.62                                         & {+4.1}                         & \cellcolor[HTML]{E6B8AF}-22.80                   & {}                     &                                                      & 644.02±3.1              & \cellcolor[HTML]{E6B8AF}+11.04 ±2.9  & 161.9                                    & 53.83 (25.8)                    \\
\textbf{ocr-img (n)}        & 2539.62±3.9           & +0.43 ±4.2                            & +0.60 ±3.8                           &                                           &                                          & 1.53                                         & -0.40                                            & -1.00                                            & {}                     &                                                      & 2491.66±0.4             & +0.68 ±0.5                           & 13.9                                     & 1.08 (0.7)      \\ 
\bottomrule              
\end{tabular}
	}
	\caption{Latency and throughput measurements showing the overheads of \lsysnop (\sys with no restoration), \lsys,\lfork, and \lfaasm relative to an unsecure baseline.
		We run 58 benchmarks across three languages indicated by \textbf{(p)} python, \textbf{(c)} c, and \textbf{(n)} \nodejs. We highlight cells of interest if there is more than a 5\% difference as follows: \protect\greenshading indicates results faorable to \sys \protect\redshading indicates results unfaorable to \sys \protect\blueshading indicates the unexpected result that \sys is outperforming the baseline. \lsys's impact on throughput can be approximated by the relative restoration time compared to the invoker's latency (inv. lat.).
		\textit{\textbf{Disclaimer}}: Faasm throughput measurements are provided for the curious reader, but no conclusions should be drawn based on them as they entangle many variables such as the difference in the platforms, their internal components and deployment, runtimes (native vs WebAssembly), as well as the isolation mechanism}
	\label{tab:latency_xput}
\end{table*}

\begin{table*}

	\centering
	\resizebox{0.98\textwidth}{!}{
		\begin{tabular}{lrrrrrrrr}
	\toprule
	\multicolumn{1}{l}{\multirow{2}{0.5cm}{\thead{Benchmark}}}&
	\multicolumn{2}{c}{\thead{baseline}}&
	\multicolumn{2}{c}{\thead{\sys}} &
	\multicolumn{4}{c}{\thead{Restoration}}\\
	\cmidrule(lr){2-3} \cmidrule(lr){4-5} \cmidrule(lr){6-9}
	& Invoker lat (ms) & T'put
	& Invoker lat (ms) & T'put
	& time (ms) & \#pages (K) & \#faults (K) & \#restored (K)\\

\midrule
cholesky (c)       &  166182.8±9208.73 &       0.02 &   175691.9±8256.49 &     0.02 &     0.57 &             0.98 &                              0.02 &                      0.01 \\
jacobi-1d (c)      &          3.8±1.25 &     671.34 &           4.2±1.64 &   578.99 &     0.62 &             0.98 &                              0.03 &                      0.02 \\
durbin (c)         &          7.6±1.35 &     314.68 &           8.0±1.48 &   295.98 &     0.62 &             0.98 &                              0.03 &                      0.02 \\
jacobi-2d (c)      &      2329.3±17.03 &       1.05 &       2343.4±14.98 &     1.05 &     0.69 &             0.98 &                              0.02 &                      0.01 \\
lu (c)             &  196555.8±11445.0 &       0.02 &  207603.5±13014.02 &     0.02 &     0.74 &             0.98 &                              0.02 &                      0.01 \\
seidel-2d (c)      &     23140.1±22.03 &       0.16 &       23139.0±21.4 &     0.16 &     0.75 &             0.98 &                              0.02 &                      0.02 \\
deriche (c)        &       1115.0±86.2 &       4.47 &       1115.0±76.95 &     4.43 &     0.75 &             0.98 &                              0.02 &                      0.01 \\
adi (c)            &    28311.1±923.24 &       0.12 &    28857.6±1215.98 &     0.12 &     0.77 &             0.98 &                              0.02 &                      0.02 \\
floyd-warshall (c) &     21151.4±39.35 &       0.17 &      21171.3±37.12 &     0.17 &     0.78 &             0.98 &                              0.02 &                      0.01 \\
bicg (c)           &         42.8±1.88 &      81.05 &          43.2±2.03 &    79.87 &     0.93 &             0.98 &                              0.03 &                      0.03 \\
fdtd-2d (c)        &      2179.1±23.85 &       0.89 &       2182.6±19.73 &     0.89 &     0.97 &             0.98 &                              0.02 &                      0.02 \\
trisolv (c)        &         23.1±1.51 &     138.18 &          23.2±2.16 &   134.92 &     0.97 &             0.98 &                              0.03 &                      0.02 \\
atax (c)           &          36.4±1.6 &      93.55 &          36.8±1.99 &    91.99 &     0.99 &             0.98 &                              0.03 &                      0.03 \\
nussinov (c)       &   39122.6±4053.11 &       0.09 &      38323.5±827.3 &     0.09 &     1.02 &             0.98 &                              0.02 &                      0.02 \\
ludcmp (c)         &  193545.9±6455.96 &       0.02 &   199550.2±8782.81 &     0.02 &     1.02 &             0.98 &                              0.03 &                      0.02 \\
mvt (c)            &        140.3±3.06 &      28.78 &          144.3±3.2 &    28.28 &     1.16 &             0.98 &                              0.04 &                      0.03 \\
doitgen (c)        &       650.5±14.61 &       5.98 &        650.0±14.79 &     5.96 &     1.31 &             0.98 &                              0.04 &                      0.02 \\
version (p)        &          3.1±1.55 &     990.38 &           4.0±1.46 &   562.89 &     1.66 &             3.14 &                              0.17 &                      0.17 \\
get-time (p)       &          2.9±1.19 &    1038.74 &            4.1±1.7 &   552.09 &     1.66 &             3.19 &                              0.18 &                      0.18 \\
covariance (c)     &     33020.6±494.9 &       0.10 &    34971.3±1084.18 &     0.10 &     1.97 &             0.98 &                              0.04 &                      0.02 \\
correlation (c)    &    32429.6±692.85 &       0.10 &    34328.9±1118.18 &     0.09 &     2.00 &             0.98 &                              0.04 &                      0.02 \\
3mm (c)            &   45729.0±1717.42 &       0.07 &    46824.4±2473.21 &     0.06 &     2.32 &             0.98 &                              0.04 &                      0.02 \\
gramschmidt (c)    &   60899.8±6020.33 &       0.06 &    64980.4±2150.99 &     0.05 &     2.53 &             0.98 &                              0.04 &                      0.02 \\
pickle (p)         &        105.6±1.89 &      35.49 &         105.7±2.11 &    34.98 &     2.90 &             3.45 &                              0.23 &                      0.23 \\
2mm (c)            &    27236.2±1544.4 &       0.12 &    28887.4±1712.35 &     0.10 &     3.12 &             0.98 &                              0.04 &                      0.02 \\
fannkuch (p)       &          4.6±1.24 &     572.32 &            6.1±2.0 &   350.22 &     3.14 &             6.12 &                              0.19 &                      0.19 \\
unpack\_seq (p)     &          3.3±1.22 &     801.86 &           5.0±2.06 &   398.15 &     3.17 &             6.12 &                              0.20 &                      0.20 \\
primes (p)         &      1829.7±53.45 &       2.04 &       1830.7±75.43 &     1.99 &     3.24 &             3.22 &                              0.51 &                      0.53 \\
json (p)           &          9.9±3.37 &     150.00 &          13.0±3.97 &   135.34 &     3.71 &             3.33 &                              0.64 &                      0.87 \\
scimark (p)        &      1812.6±30.71 &       2.12 &       1806.6±28.47 &     2.12 &     3.77 &             3.26 &                              0.51 &                      0.52 \\
telco (p)          &         155.6±3.8 &      25.01 &         158.0±2.95 &    23.77 &     3.91 &             3.29 &                              0.53 &                      0.53 \\
json\_loads (p)     &        102.0±1.95 &      36.46 &         103.3±2.14 &    35.29 &     4.04 &             6.12 &                              0.22 &                      0.22 \\
nbody (p)          &       2823.7±69.0 &       1.34 &       2845.0±53.46 &     1.34 &     4.08 &             6.12 &                              0.21 &                      0.21 \\
richards (p)       &        353.1±4.64 &      10.68 &         351.1±4.41 &    10.85 &     4.16 &             6.18 &                              0.23 &                      0.23 \\
md2html (p)        &         31.0±1.95 &      93.94 &          32.7±2.31 &    88.50 &     4.25 &             4.93 &                              0.63 &                      0.62 \\
spectral (p)       &        592.8±9.92 &       6.45 &        605.2±14.14 &     6.40 &     4.29 &             6.12 &                              0.22 &                      0.21 \\
hexiom (p)         &        218.2±4.21 &      17.45 &         219.2±3.98 &    17.28 &     4.35 &             6.18 &                              0.28 &                      0.28 \\
raytrace (p)       &      2459.2±67.26 &       1.58 &       2463.9±51.19 &     1.57 &     4.42 &             6.25 &                              0.36 &                      0.35 \\
deltablue (p)      &         20.4±1.61 &     157.63 &          21.3±2.02 &   140.26 &     4.64 &             6.18 &                              0.23 &                      0.33 \\
logging (p)        &     1249.4±652.55 &       0.00 &          227.9±5.2 &    16.34 &     4.77 &             6.12 &                              0.42 &                      0.41 \\
json\_dumps (p)     &         533.1±6.0 &       7.19 &         551.5±9.92 &     6.95 &     4.92 &             6.37 &                              0.51 &                      0.51 \\
chaos (p)          &       648.5±86.06 &       6.03 &        652.0±39.21 &     5.94 &     4.93 &             6.32 &                              0.47 &                      0.47 \\
float (p)          &         27.1±1.92 &     125.98 &          27.8±1.87 &   109.09 &     4.99 &             6.26 &                              0.65 &                      0.65 \\
pidigits (p)       &       2347.6±5.76 &       1.64 &        2349.1±6.51 &     1.63 &     5.40 &             6.14 &                              0.81 &                      0.81 \\
sentiment (p)      &          6.5±1.76 &     385.07 &           8.9±3.18 &   230.39 &     6.00 &            16.86 &                              0.57 &                      0.57 \\
pyaes (p)          &      4672.0±63.68 &       0.82 &       4751.3±61.36 &     0.80 &     6.02 &             6.21 &                              0.83 &                      0.84 \\
go (p)             &        593.0±6.64 &       6.48 &         596.6±5.69 &     6.42 &     6.90 &             6.25 &                              0.84 &                      0.95 \\
base64 (p)         &        743.2±7.11 &       5.18 &        761.5±10.48 &     5.10 &     7.67 &             5.13 &                              1.86 &                      1.66 \\
mdp (p)            &      6345.5±63.96 &       0.59 &       6412.3±82.13 &     0.58 &     9.55 &             7.33 &                              2.22 &                      2.85 \\
pyflate (p)        &      1599.8±16.39 &       2.39 &       1622.5±12.58 &     2.34 &    11.67 &             8.25 &                              3.01 &                      2.33 \\
get-time (n)       &          3.7±1.29 &     942.07 &           6.4±3.58 &   133.45 &    12.58 &           156.76 &                              0.59 &                      0.64 \\
json (n)           &          9.4±3.55 &     159.09 &          16.1±4.94 &    86.58 &    13.02 &           156.78 &                              0.67 &                      0.85 \\
autocomplete (n)   &          3.8±1.41 &     922.59 &           6.3±3.41 &   121.98 &    13.52 &           156.98 &                              0.69 &                      0.92 \\
ocr-img (n)        &      2491.7±10.63 &       1.53 &       2508.5±12.24 &     1.52 &    13.95 &           156.80 &                              0.89 &                      1.08 \\
heat-3d (c)        &      3059.5±81.59 &       1.02 &       3272.0±28.01 &     0.98 &    16.09 &             4.35 &                              0.02 &                      3.39 \\
img-resize (n)     &       445.3±74.34 &       6.57 &       721.7±110.76 &     4.10 &    61.83 &           179.43 &                              9.58 &                     18.05 \\
primes (n)         &       274.6±20.11 &      11.79 &         287.1±23.1 &     8.16 &    84.74 &           201.35 &                              1.27 &                     34.20 \\
base64 (n)         &       644.0±20.22 &       5.62 &        715.1±20.89 &     4.34 &   161.93 &           208.42 &                             47.98 &                     53.83 \\
\bottomrule
\end{tabular}

	}
	\caption{\sys's overhead on the throughput is a function of \sys's added overhead both on the critical path (\#faults) which has negligible overhead on most functions, as well as the overhead off the critical path  (Restoration time) which is mostly a function of the address space size (\#pages (K)) and the number of restored pages (\#restored (K)) as well as restoring the memory layout as demonstrated earlier in ~\autoref{fig:restoration_overheads}. Data is sorted by the restoration time.}
	\label{tab:small-table}
\end{table*}

\end{document}